\documentclass[iop]{emulateapj}
\usepackage{color}
\usepackage{amsmath}
\usepackage{natbib}
\usepackage[utf8]{inputenc}
\definecolor{lila}{rgb}{0.4,0,1}
\definecolor{grau}{rgb}{0.5,0.5,0.5}
\definecolor{darkblack}{rgb}{0.2,0.2,1}
\definecolor{orange}{rgb}{1,0.5,0}

\newcommand{\MJ}{M_{\rm J}}
\newcommand{\RJ}{R_{\rm J}}

\shorttitle{\textit{Ab initio} equations of state for hydrogen and helium}
\shortauthors{Becker et al.}

\begin{document}

\title{\textit{Ab initio} equations of state for hydrogen (H-REOS.3) and helium (He-REOS.3) and their implications for the interior of Brown Dwarfs}

\author{Andreas Becker$^1$, Winfried Lorenzen$^1$, Jonathan J.\ Fortney$^{2}$, Nadine Nettelmann$^2$, Manuel Sch\"ottler$^1$,
 and Ronald Redmer$^1$}
\affil{$^1$Institut f\"ur Physik, Universit\"at Rostock, D-18051 Rostock, Germany\\
$^2$Department of Astronomy \& Astrophysics, University of California, Santa Cruz, CA 95064, USA
}

\begin{abstract}
We present new equations of state (EOS) for hydrogen and helium covering a wide range of temperatures from 60~K to 10$^7$~K and densities 
from $10^{-10}$~g/cm$^3$ to $10^3$~g/cm$^3$. They include an extended set of \textit{ab initio} EOS data for the strongly correlated quantum regime with an accurate
connection to data derived from other approaches for the neighboring regions. We compare linear-mixing isotherms based on our EOS tables with available real-mixture data. A first important 
astrophysical application of this new EOS data is the calculation of interior models for Jupiter and the comparison with recent results. 
Secondly, mass-radius relations are calculated for Brown Dwarfs which we compare with predictions derived from the widely used EOS of 
Saumon, Chabrier and van Horn. Furthermore, we calculate interior models for typical Brown Dwarfs with different masses, namely Corot-3b, 
Gliese-229b and Corot-15b, and the Giant Planet KOI-889b. The predictions for the central pressures and densities differ by up to 10$\%$ dependent 
on the EOS used. Our EOS tables are made available in the supplemental material of this paper.

\end{abstract}

\keywords{equation of state -- dense matter -- plasmas -- stars: low-mass, brown dwarfs -- planets and satellites: individual(Jupiter)}

\section{Introduction}\label{sec:intro}

Hydrogen and helium are the most abundant elements in the universe. The knowledge of their equations of state is of fundamental interest for modeling 
astrophysical objects such as stars, Brown Dwarfs (BDs), and Giant Planets (GPs). As discussed by \cite{Stevenson1991}, observations and interior modeling 
of BDs can test the EOS of non-ideal degenerate matter. Furthermore, recent high-pressure experiments at the National Ignition Facility have reached the 
multi-Gigabar regime~\citep{Hurricane2014} so that states deep in Brown Dwarfs and the corresponding equation of state (EOS) can now 
be probed via isentropic compression experiments.

The current map of theoretical EOS data can be divided into chemical models which often cover a large area in the $T-\rho$ or $T-P$ plane, and EOS data derived from \textit{ab initio} simulations at individual $T-\rho$ points.
The great advantage of chemical models is that they provide EOS data from the classical ideal gas limit up to degenerate matter via a single free energy model. 
A widely used EOS for modeling GPs and BDs within this approach is that of Saumon, Chabrier and van Horn (SCvH-EOS)~\citep{Saumon1995}.
An inherent problem of chemical models is, however, the treatment of correlations between the various species via effective pair potentials and the choice of appropriate reference systems. 
This is crucial for conditions where pressure dissociation and ionization occur in hydrogen and helium. For instance, the maximum compression of hydrogen along the 
principle Hugoniot curve is about 4.25-4.5 as derived from shock-wave experiments~\citep{Knudson2004,Sano2011}, while the SCvH model predicts a considerably higher 
ratio of $\sim5.5$, see Fig.~3 in~\cite{Loubeyre2012}. Therefore, chemical models EOS are of limited accuracy in that strongly correlated quantum regime. 

On the other hand, \textit{ab initio} simulations performed for hydrogen and helium (for a recent review, see~\cite{McMahon2012}) yield very good agreement with experimental data (for hydrogen, see \cite{Becker2013} and references therein). However, first principles simulations are numerically expensive and can only be performed for a limited grid of $\rho-T$ points. Although a free energy can be fitted to the calculated data set afterwards, see \cite{Militzer2009,Morales2010,Caillabet2011}, its application is restricted to that specific region of the $\rho-T$ plane. There is no \textit{ab initio} method available that can generate EOS data from the classical ideal gas up to the degenerate limit for all desired temperatures. 

The purpose of this paper is to present EOS data for hydrogen and helium that cover the wide range of densities and temperatures as typical for chemical models, but in addition have the accuracy of \textit{ab initio} data. Two main problems have to be solved in order to reach that goal. Firstly, extended \textit{ab initio} simulations were performed for the strongly correlated quantum region within the framework of density functional theory molecular dynamics (DFT-MD). Secondly, this accurate EOS data set has to be connected with EOS data of similar accuracy that are valid in the neighboring regions of the $\rho-T$ plane. We have tested various chemical and \textit{ab initio} approaches and checked thermodynamic consistency to a large extent. The final data tables representing a substantially improved EOS for both hydrogen and helium are made available as online supplemental material.

Although a couple of EOS data sets exist for hydrogen that are derived from first principles simulations \citep{Caillabet2011,Hu2011,Wang2013,Morales2012}, these are mostly dedicated to applications in inertial confinement fusion (ICF). To our knowledge, the current work is the first project that covers the Brown Dwarf regime using \textit{ab initio} EOS data for hydrogen and helium.

Having the new EOS data at our disposal, we then calculate interior models for Jupiter as the prototypical GP in our solar system and for various Brown Dwarfs as well as mass-radius relations for Brown Dwarfs. We compare with results derived using the SCvH-EOS.

Our paper is arranged as follows. We present and describe in detail our EOS data with the focus on the new helium EOS in Sec.~\ref{sec:HEOS}. 
In particular, we compare experimental principal Hugoniot curves for helium with predictions from our new helium EOS in Sec.~\ref{subsec:Experiments}.
Furthermore, we investigate the difference of a linear mixing EOS composed of our data and a real mixture EOS \citep{Militzer2013b} in 
Sec.~\ref{subsec:LMvsRM}. We then apply the EOS data to Jupiter models in Sec.~\ref{sec:Jupi} and to mass radius relations and interior models 
of selected Brown Dwarfs in Sec.~\ref{sec:IntBrownie}, finishing with the conclusions in Sec.~\ref{sec:Conclusion}.

\section{The equations of state for hydrogen and helium}\label{sec:HEOS}
Before we describe our current EOS tables, we give a brief summary of the previous hydrogen and helium EOS data developed by our group. These data led to interior models of Jupiter published in \cite{Nettelmann2008} and \cite{Nettelmann2012}. The only He-REOS (helium Rostock EOS) to date has been composed from the Sesame 5761 data \citep{Lyon1992} and DFT-MD data derived from simulations with 32-64 particles \citep{Kietzmann2007} for 4 isotherms (4000~K, 6310~K, 15850~K, 31620~K) and densities between 0.16 and 10~g/cm$^3$, see \cite{Nettelmann2008}.

The first EOS for hydrogen (H-REOS.1) already contained an extended data set of DFT-MD data derived from simulations with 64 particles \citep{Holst2008} and was applied in \cite{Nettelmann2008}. 
The DFT-MD data has been connected to FVT and FVT+ EOS data (see below) within H-REOS.1 and H-REOS.2. The latter contained an extended DFT-MD data set as well but derived from fully converged simulations with 256 particles.
This improvement led to Jupiter models fulfilling all observational constrains \citep{Nettelmann2012}. Obviously, a better knowledge of the EOS is linked to progress in the field of \textit{ab initio} simulations.
Particle numbers of 256 for hydrogen and 108 for helium as used throughout this paper could be achieved due to increasing computational power.

\begin{figure}
 \plotone{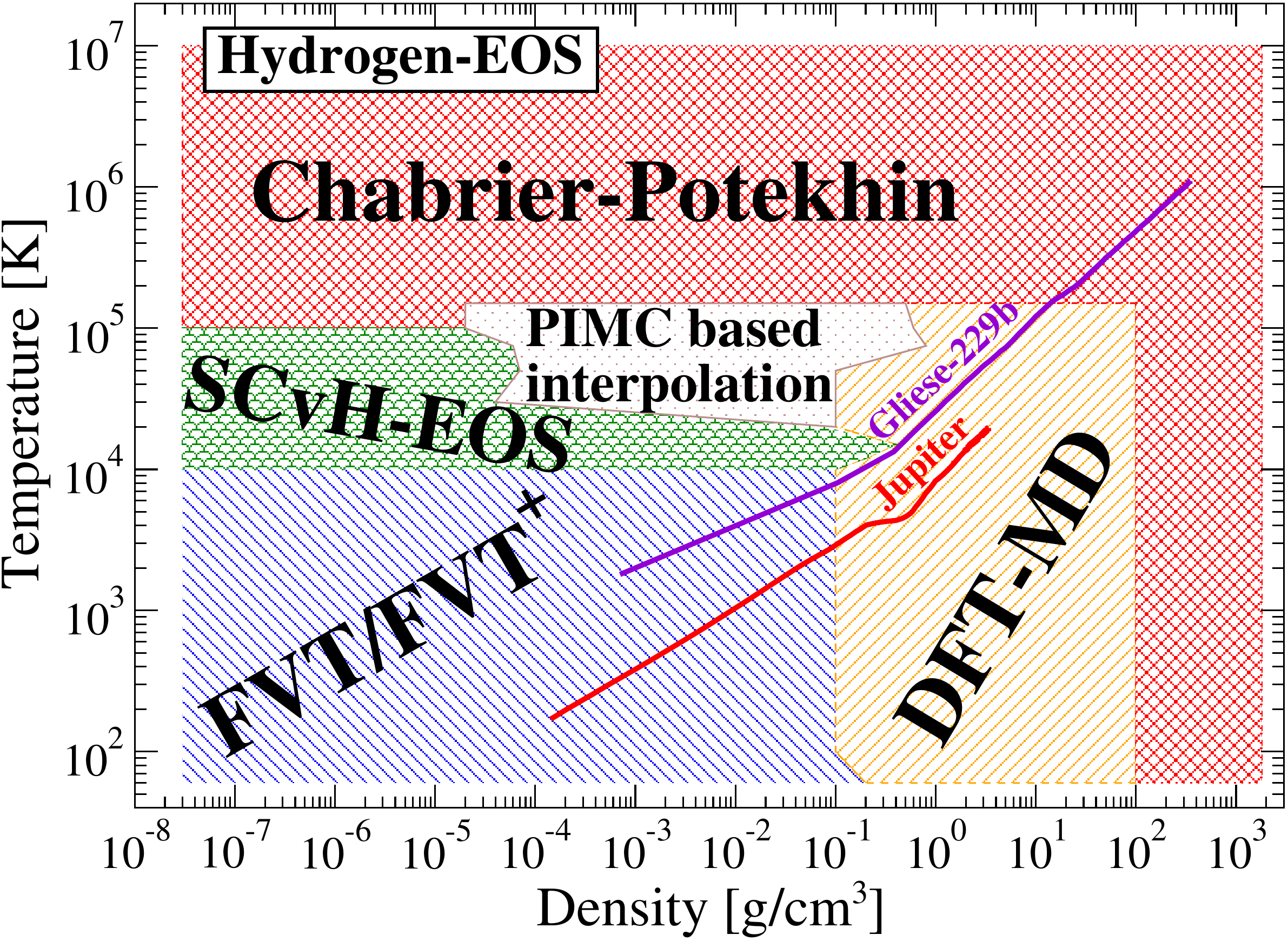}
 \caption{\label{fig:HArea}Constituents of the hydrogen EOS table (H-REOS.3). The red curve represents the Jovian adiabat and the violet 
 one the adiabat of the Brown Dwarf Gliese-229b with respect to the partial density of hydrogen in the mixture EOS, see Sec.~\ref{subsec:LMvsRM}.
 }
 \end{figure}

Our final EOS tables are composed of different models, see Fig.~\ref{fig:HArea} for hydrogen and Fig.~\ref{fig:HeArea} for helium. A compact compilation of these models is summarized in Tab.~\ref{tab:models}.
The centerpiece of the EOS is the DFT-MD data for the strongly correlated quantum region in both cases. Proper connections to the neighboring $\rho-T$ regions 
have been performed. Since our DFT-MD data for hydrogen and the respective EOS for planetary modeling (see the Jovian adiabat in Fig.~\ref{fig:HArea}) have been 
published elsewhere \citep{Nettelmann2012,Becker2013} we will focus here on interior models for BDs. For this purpose, the DFT-MD data set used in H-REOS.2 has been 
extended considerably toward high temperatures and densities up to 70~g/cm$^3$ (H-REOS.3). However, the main part of this section is dedicated to the new and so far 
unpublished helium EOS (He-REOS.3).

\begin{deluxetable*}{lccc}
\tabletypesize{\tiny}
\tablecolumns{4}
\tablecaption{\label{tab:models}Compilation of the equation of state (EOS) models used in this paper.}
\tablehead{
\colhead{Abbreviation} & \colhead{Full name}   & \colhead{Features of the EOS model} & \colhead{References} }
\startdata
&Density-functional theory&\textit{Ab initio} calculations using VASP, correlations and&\cite{Kresse1993}\\
DFT-MD &  molecular dynamics simulations  &quantum effects are treated consistently&\cite{Kresse1994}\\
&at finite temperatures&within DFT&\cite{Kresse1996}\\
\hline
&Hydrogen-&EOS for GP range, combination of FVT, FVT+&\\
H-REOS.1&Rostock equation of state & and DFT-MD data derived from&\cite{Nettelmann2008}\\
&v1&simulations with 64 particles&\\
\hline
&Hydrogen-&EOS for GP range, combination of FVT, FVT+&\\
H-REOS.2&Rostock equation of state&and DFT-MD data derived from&\cite{Nettelmann2012}\\
&v2&simulations with 256 particles&\\
\hline
&Hydrogen-&EOS for GP and BD range, combination of FVT,&\\
H-REOS.3&Rostock equation of state& FVT+, SCvH, PIMC and DFT-MD data &this paper\\
&v3&derived from simulations with 256 particles&\\
\hline
&Helium-&EOS for GP range, combination of Sesame 5761&\\
He-REOS.1&Rostock equation of state&and DFT-MD data for four isotherms&\cite{Nettelmann2008}\\
& v1&derived from simulations with 32/64 particles&\\
\hline
&Helium-&EOS for GP and BD range, combination of SCvH,&\\
He-REOS.3&Rostock equation of state&a virial EOS and DFT-MD data &this paper\\
& v3&derived from simulations with 108 particles&\\
\hline
&Saumon, Chabrier and van Horn&Chemical model EOS for GP and BD range, minimization&\\
SCvH&equations of state for H and He&of free energy using fluid perturbation theory&\cite{Saumon1995}\\
\hline
&Fluid variational theory&Chemical model EOS, minimization of free energy using&\\
FVT&hydrogen equation of state&fluid variational theory &\cite{Juranek2002}\\
\hline
&Fluid variational theory ``plus''&Chemical model EOS based on FVT data with&\\
FVT+&hydrogen equation of state&additional treatment of ionization&\cite{Holst2007}\\
\hline
&Chabrier-Potekhin model&Free energy model for fully&\\
CP&for the hydrogen equation of state&ionized plasma at arbitrary degeneracy&\cite{Chabrier1998}\\
\hline
&Path-integral Monte Carlo&\textit{Ab initio} method based on the evaluation &\\
PIMC&simulations& of the N-body density matrix&\cite{Hu2011}
\enddata
\end{deluxetable*}

The REOS.3 tables are arranged in four columns according to isotherms. The first column contains the density $\rho$ in g/cm$^3$, followed by the temperature T in K, 
the pressure P in GPa and the specific internal energy u in kJ/g, see Tabs.~\ref{tab:H-REOS} and ~\ref{tab:He-REOS} which can be obtained in a machine-readable form from the online version of this paper.
Note that our current tables do not contain entropy values because they are not directly accessible via DFT-MD simulations. Their calculations via the power spectrum as proposed by \cite{Desjarlais2013} or using thermodynamic integration as performed, e.g., by \cite{Militzer2013b} and \cite{Morales2013a} will be done in the future. However, isentropic paths as needed for the interiors of Giant Planets and Brown Dwarfs can be obtained solely by the knowledge of the pressure and the internal energy via an integration scheme \citep{Nettelmann2012} or a differential equation \citep{Becker2013}.

Due to the lack of entropies we cannot construct the free energy and derive all thermodynamic values from that canonical potential. Accordingly, we spliced together the different EOS data in pressure and internal energy simultaneously with smooth transitions, conserving thermodynamic consistency to a large extent. The details are given in the following sections.
\begin{deluxetable}{cccc}
 \tablewidth{0pc}
 \tabletypesize{\small}
 \tablecaption{\label{tab:H-REOS}Short example of the H-REOS.3 table }
 \tablehead{ \colhead{$\rho$ [g/cm$^3$]} & \colhead{T [K]}& \colhead{P [GPa]} & \colhead{u [kJ/g]}  }
 \startdata
 $3.0000\cdot10^{-8}$ & $6.0000\cdot10^{1}$ & $7.4249\cdot10^{-9}$ & $-1.5895\cdot10^{3}$ \\
 \vdots & \vdots & \vdots & \vdots  \\
 $1.8000\cdot10^{3}$ & $6.0000\cdot10^{1}$ & $2.4854\cdot10^{8}$ & $1.9104\cdot10^{5}$ \\
 $3.0000\cdot10^{-8}$ & $1.0000\cdot10^{2}$ & $1.2375\cdot10^{-8}$ & $-1.5891\cdot10^{3}$ \\
 \vdots & \vdots & \vdots & \vdots  \\
 $1.8000\cdot10^{3}$ & $1.0000\cdot10^{1}$ & $2.4855\cdot10^{8}$ & $1.9105\cdot10^{5}$ \\
 \vdots & \vdots & \vdots & \vdots  \\
 $3.0000\cdot10^{-8}$ & $1.0000\cdot10^{7}$ & $4.9887\cdot10^{-3}$ & $2.4943\cdot10^{5}$ \\
 \vdots & \vdots & \vdots & \vdots  \\
 $1.8000\cdot10^{3}$ & $1.0000\cdot10^{7}$ & $4.5435\cdot10^{8}$ & $3.6763\cdot10^{5}$ 
 \enddata
 \tablecomments{This table is published in its entirety in the electronic edition of the Astrophysical Journal.
A portion is shown here for guidance regarding its form and content.}
 \end{deluxetable}
 
 \begin{deluxetable}{cccc}
 \tablewidth{0pc}
 \tabletypesize{\small}
 \tablecaption{\label{tab:He-REOS} Short example of the He-REOS.3 table }
 \tablehead{ \colhead{$\rho$ [g/cm$^3$]} & \colhead{T [K]}& \colhead{P [GPa]} & \colhead{u [kJ/g]}  }
 \startdata
 $1.0000\cdot10^{-10}$ & $6.0000\cdot10^{1}$ & $1.2463\cdot10^{-11}$ & $-1.8429\cdot10^{3}$ \\
 \vdots & \vdots & \vdots & \vdots  \\
 $1.0000\cdot10^{3}$ & $6.0000\cdot10^{1}$ & $2.69412\cdot10^{7}$ & $3.2974\cdot10^{4}$ \\
 $1.0000\cdot10^{-10}$ & $1.0000\cdot10^{2}$ & $2.0773\cdot10^{-11}$ & $-1.8428\cdot10^{3}$ \\
 \vdots & \vdots & \vdots & \vdots  \\
 $1.0000\cdot10^{3}$ & $1.0000\cdot10^{2}$ & $2.69413\cdot10^{7}$ & $3.2974\cdot10^{4}$ \\
 \vdots & \vdots & \vdots & \vdots  \\
 $1.0000\cdot10^{-10}$ & $1.0000\cdot10^{7}$ & $6.2318\cdot10^{-6}$ & $9.3538\cdot10^{4}$ \\
 \vdots & \vdots & \vdots & \vdots  \\
 $1.0000\cdot10^{3}$ & $1.0000\cdot10^{7}$ & $7.1984\cdot10^{8}$ & $1.0273\cdot10^{5}$ 
 \enddata
 \tablecomments{This table is published in its entirety in the electronic edition of the Astrophysical Journal.
A portion is shown here for guidance regarding its form and content.}
 \end{deluxetable}

\subsection{Hydrogen}\label{subsec:Hydrogen}
\subsubsection{DFT-MD data}

The DFT-MD framework combines classical molecular dynamics simulations for the ions 
with a quantum treatment for the electrons based on DFT for finite temperatures~\citep{Mermin1965} 
which is implemented in the VASP code~\citep{Kresse1993, Kresse1994, Kresse1996}. 
The Coulomb interactions between the electrons and ions are treated using 
projector-augmented wave (PAW) potentials~\citep{Blochl1994, Kresse1999} 
at densities below 9~g/cm$^3$ with a converged energy cutoff of 1200~eV. 
For densities between $9\leq\rho\leq70$~g/cm$^3$ we applied the full Coulomb potential with a significantly higher energy cutoff of 3-10~keV.
The ion temperature is controlled with a Nos\'e thermostat~\citep{Nose1984}.
A detailed description of the zero point motion treatment of the ions bound in molecules can be found in \cite{Becker2013}.

Convergence of the results is checked with respect to the particle number, the 
$\mathbf{k}$-point sets used for the evaluation of the Brillouin zone, 
and the energy cutoff for the plane wave basis set. For the simulations 
we chose 256~atoms and the Baldereschi mean value point~\citep{Baldereschi1973} 
which proved to yield well converged results~\citep{Holst2008, Lorenzen2010}.

We use the exchange-correlation functional of Perdew, Burke, and Ernzerhof (PBE)~\citep{Perdew1996} which has been shown to 
give precise results for warm dense matter states~\citep{Desjarlais2003,Lorenzen2010,Holst2008}.

\subsubsection{EOS table composition}

The several components of our hydrogen EOS table (H-REOS.3) are shown in Fig.~\ref{fig:HArea}. The orange region covers states of matter that represent the greatest challenge for many-particle theory since strong correlations and quantum effects occur that lead to pressure-induced dissociation and ionization processes. 
An appropriate method to describe such states is the DFT-MD technique, see previous subsection.

\begin{figure}
   \plotone{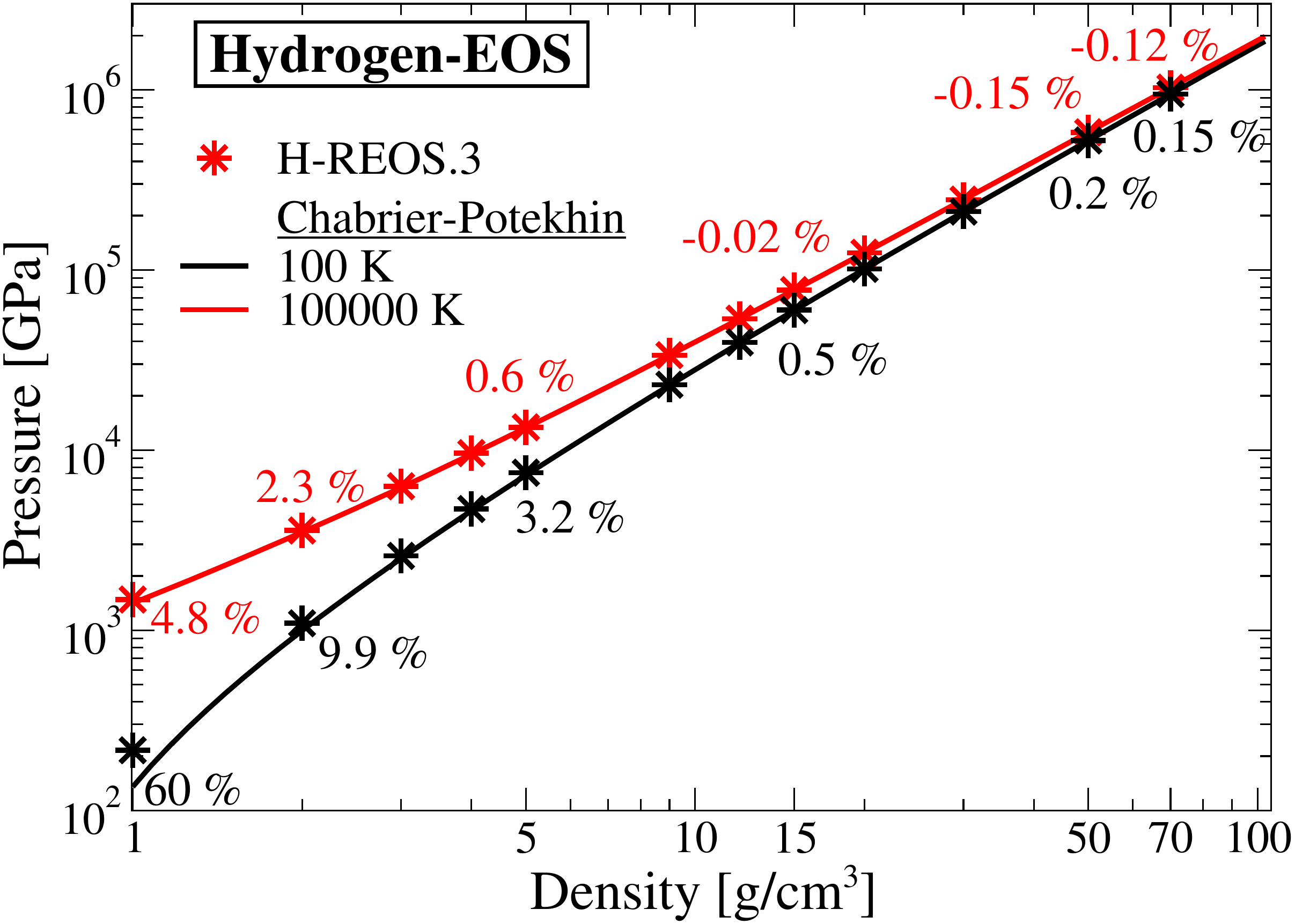}
   \caption{\label{fg:HighDense}Connection of DFT-MD data (stars) with the Chabrier-Potekhin EOS (solid lines) in the high-density limit; relative deviations are given at selected data points. }
\end{figure}

To cover the remaining parts of the $\rho$-T plane we connected the DFT-MD data to EOS data derived from chemical models such as the fluid variational theory (FVT) for the molecular and dissociated fluid, see~\cite{Juranek2002}, the SCvH-EOS \citep{Saumon1995} for the partially ionized plasma at lower densities, and the Chabrier-Potekhin (CP) model for fully ionized matter~\citep{Chabrier1998}.
Note that interpolations between two EOS models had only to be performed at the interface between the DFT-MD data and the region labeled with ``PIMC based interpolation'' in Fig.~\ref{fig:HArea}, see also below. At all remaining interfaces the points of the different EOS models merge smoothly into each other.
In particular, the CP model connects almost perfectly to the DFT-MD data at high densities, see Fig.~\ref{fg:HighDense}. For low temperatures (black: 100~K isotherm) as well as for high temperatures (red: 100~kK isotherm) the transition at 70~g/cm$^3$ is performed with a deviation $\leq 0.15\%$ in the thermal EOS $P(\rho,T)$. We find the same accuracy for intermediate temperatures in $P(\rho,T)$ as well as for the transition within the caloric EOS $u(\rho,T)$.

\begin{figure}
   \plotone{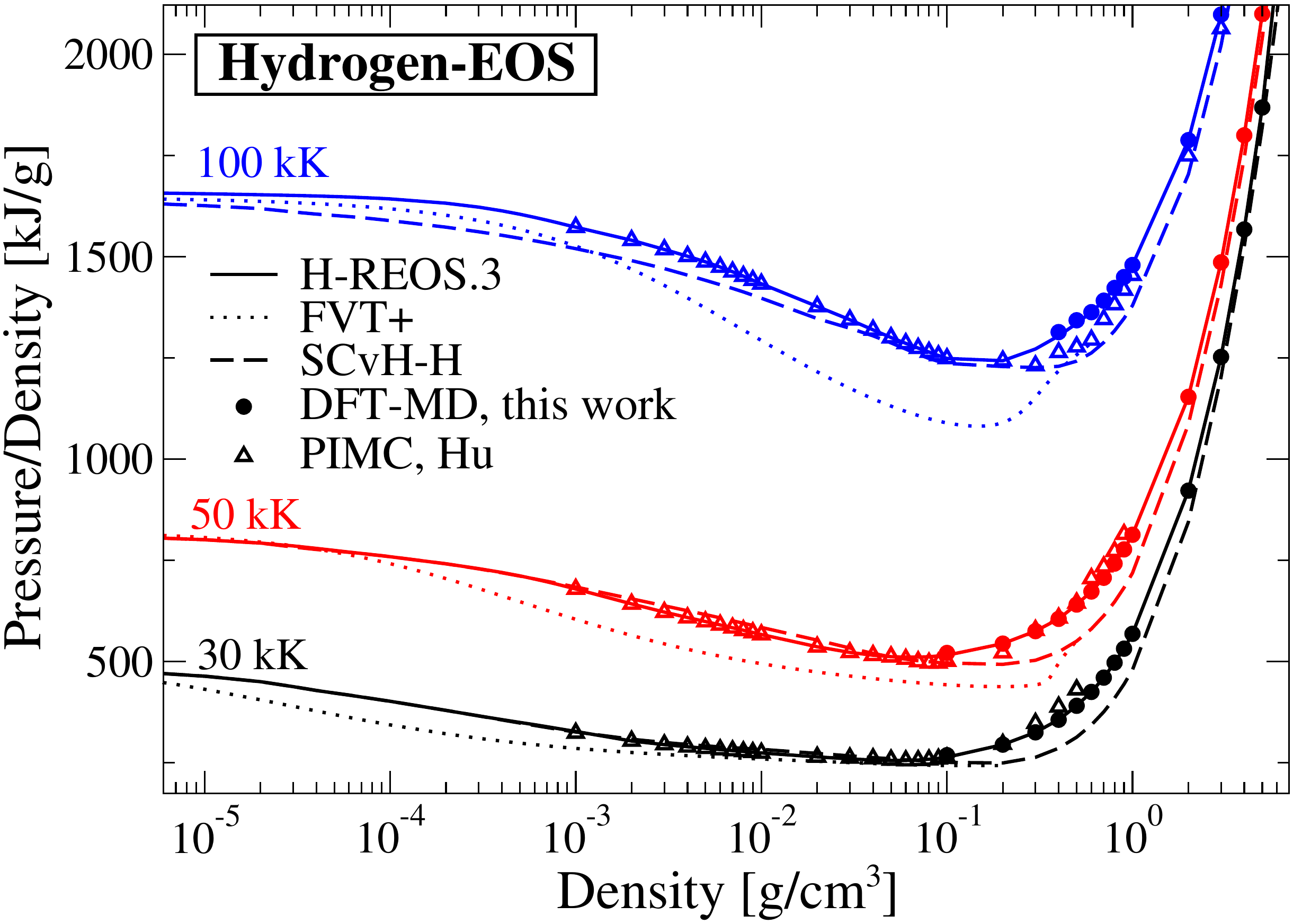}
   \caption{\label{fg:Hinterpol}Interpolation region of the hydrogen EOS represented by the 30~kK (black), 50~kK (red) and 100~kK (blue) isotherms for the different EOS models. Dashed curves are the SCvH data, dotted lines are FVT+ results, and solid lines are the final H-REOS.3 isotherms containing SCvH data, PIMC results from Hu (open triangles) and DFT-MD data (circles).}
\end{figure}

However, the low-density connection to the DFT-MD data turned out to be more difficult. Up to 10~kK we used the FVT data which lead to similar results as the SCvH-EOS but can be connected more smoothly to the DFT-MD data. The subtle details appear in regions where significant ionization occurs, e.g, in the transition region left of the DFT-MD data above 10~kK, see Fig.~\ref{fig:HArea}. The choice between the FVT+ model (FVT including ionization, see \cite{Holst2007}) and the SCvH model came out in favor of the latter for the following reason: We compared the path integral Monte Carlo (PIMC) results from \cite{Hu2011} with the 30~kK, 50~kK, and 100~kK isotherms of the FVT+ and the SCvH-EOS, see Fig.~\ref{fg:Hinterpol}.
We plot the ratio of pressure and density versus the density. In this $P/\rho$-representation, non-ideal contributions to the EOS can nicely be seen as deviations from horizontal lines that represent the ideal gas law ($P/\rho=RT/M$ with the molar mass $M$).

It turns out that the accurate PIMC data (triangles) clearly favor a connection from the SCvH data (dashed lines) to the DFT-MD data (filled circles); the FVT+ data (dotted lines) underestimate the pressure substantially. For 30~kK and 50~kK the PIMC data are very close to the SCvH results in the low-density range. When the SCvH data start to underestimate the pressures at densities above 0.1~g/cm$^3$, the PIMC results turn into the direction of the DFT-MD data -- both \textit{ab initio} methods agree there nicely. 
Thus, we have performed a smooth interpolation from the SCvH data via the PIMC data to the DFT-MD results for the final H-REOS.3 isotherms (solid lines) from 20~kK to 150~kK, see interpolation regime in Fig.~\ref{fig:HArea}.
In detail, the following data points of the H-REOS.3 table are generated using natural cubic spline interpolations: 0.05-0.1~g/cm$^3$ at 20~kK, 0.03-0.1~g/cm$^3$ at 30~kK, 0.06-0.2~g/cm$^3$ at 50~kK, 0.08-0.2~g/cm$^3$ at 75~kK, 0.3-0.5~g/cm$^3$ at 100~kK, and 0.1-0.5~g/cm$^3$ at 150~kK.
For temperatures above 150~kK hydrogen is fully ionized and well described by the CP model. This can be connected to the SCvH data at lower densities without violating thermodynamic consistency, see Sec.~\ref{subsec:TdynCons}.

\subsection{Helium}\label{subsec:Helium}

\begin{figure}
 \plotone{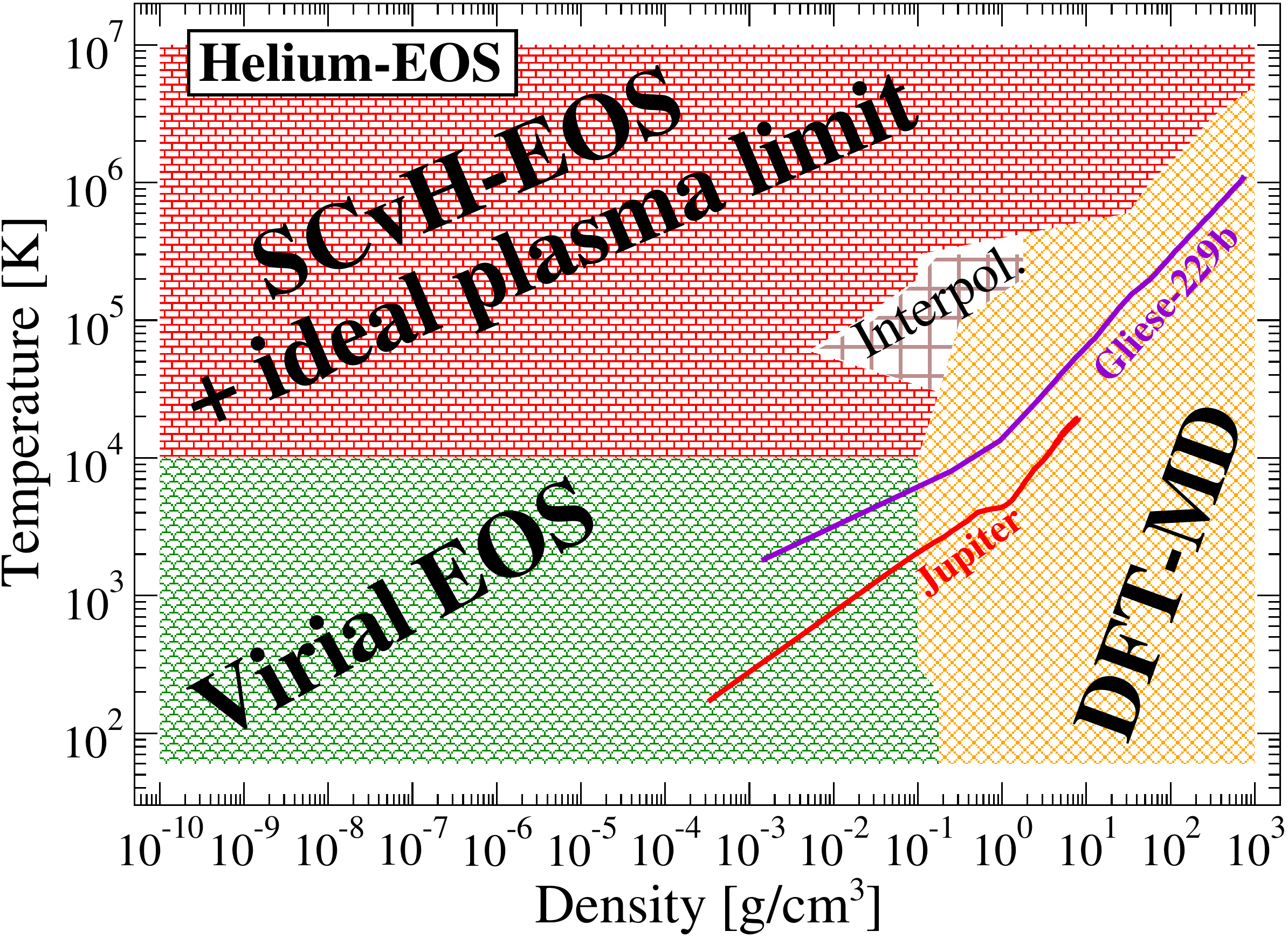}
 \caption{\label{fig:HeArea}Constituents of the helium EOS. A separate interpolation between the SCvH-He EOS data and the DFT-MD data was neccessary for low densities between 60~kK and 300~kK. The red curve represents the Jovian adiabat and the violet one the adiabat of the Brown Dwarf Gliese-229b with respect to the partial density of helium in the mixture EOS, see Sec.~\ref{subsec:LMvsRM}.}
 \end{figure}

The helium EOS (He-REOS.3) consists of three major parts, see Fig.~\ref{fig:HeArea}: DFT-MD EOS data cover the intermediate and high-density range of the density-temperature plane, where correlation and degeneracy effects are important and have to be treated accurately, while for temperatures up to 10~kK and low densities we use a virial EOS based on an ideal gas model and very accurate virial coefficients, see Sec.~\ref{Sec:Virial}. It turned out that the helium EOS model of \cite{Saumon1995} (SCvH-He) connects smoothly to our DFT-MD data for most $T-\rho$ points and is, thus, the most appropriate one for our purpose.
Interpolations were only performed between 60~kK and 300~kK since ionization of the He atoms occurs in the transition region between the SCvH and DFT-MD data. In the following we describe the methods used for generating the EOS data, discuss the composition of the final EOS table, and compare with other \textit{ab initio} data for helium.

\subsubsection{DFT-MD data}

The DFT-MD EOS data for helium are calculated using the VASP code with 108 particles (216 electrons) in the simulation box. As for hydrogen, we performed the $\mathbf{k}$-point sampling using the Baldereschi mean value point, applied the Nos$\acute{\mathrm{e}}$ thermostat to control the ion temperature and the PBE exchange-correlation functional. The used potentials and the energy-cutoff differ with respect to the simulated temperatures and densities. For $\rho\leq 10$~g/cm$^3$ we applied the PAW-potential. We used a cutoff of 1300-1400~eV for densities below 1~g/cm$^3$ and temperatures below 10~kK. The density range between 1 and 10~g/cm$^3$ for temperatures below 10~kK was evaluated with 800~eV, that was also used for all densities $\rho\leq 10$~g/cm$^3$ and temperatures above 10~kK.

For densities above 10~g/cm$^3$ and all temperatures we applied again the full Coulomb potential with a cutoff of 10~keV for well converged results. The respective phase diagram of helium derived from this data can be found elsewhere~\citep{Lorenzen2012}. DFT-MD simulations with VASP become more demanding at higher temperatures because the electrons occupy increasingly higher energy bands which have to be taken into account. On the other hand, the number of bands decreases with increasing density. That is the reason why simulations of very high temperatures are only possible at very high densities. However, for high temperatures and/or lower densities, the system becomes increasingly ideal and can be described appropriately by other methods; these are presented in the following subsection.

\subsubsection{The virial and high-temperature EOS}\label{Sec:Virial}

For the intermediate region between the strongly correlated quantum regime treated with DFT-MD and the ideal gas limit we introduced correction terms to the ideal gas equation. They describe weak coupling between the particles and result in a smooth transition between the DFT-MD data and the ideal gas EOS. Such correction terms can be expressed by virial coefficients $B_i(T)$. The respective virial EOS is composed of an ideal part $P^{\mathrm{id}}$ and a correlation part $P^{\mathrm{cor}}$ as follows:
\begin{equation}
 P(\rho,T) = P^{\mathrm{id}} + P^{\mathrm{cor}} = \frac{\rho RT}{M} \left[ 1 + 
 \sum\limits_{i=2}^\infty B_i(T)\left( \frac{\rho}{M}\right)^{i-1}\right] \;.
 \label{eq:Virial}
\end{equation}
$R$ is the universal gas constant and $M$ the molar mass of helium. Note that in this notation the ideal part $\rho RT/M$ is without ionization, while we finally use a more elaborated ideal gas model with ionization, see below.
The contribution of the virial coefficients to the specific internal energy $u=U/m$ is calculated via the fundamental connection between the thermal and caloric EOS,
\begin{equation}
 -\rho^2\left(\frac{\partial u}{\partial \rho}\right)_T = T\left(\frac{\partial P}{\partial T}\right)_{\rho} -P \quad.
 \label{eq:fundamental}
\end{equation}
Using Eqs.~(\ref{eq:Virial}) and (\ref{eq:fundamental}) we obtain:
\begin{equation}
 u(\rho,T) = u^{\mathrm{id}}-\frac{RT^2}{M}\left[ \sum \limits_{i=2}^\infty   \frac{\mathrm{d}B_i(T)}{\mathrm{d}T}\frac{1}{i-1}\left(\frac{\rho}{M}\right)^{i-1}  \right] \,.
\end{equation}
We investigated several sets of virial coefficients from $B_2(T)$ up to $B_5(T)$ \citep{Slaman1991, Bich2007, Cencek2012,Shaul2012a, Shaul2012} to find the best transition from the ideal gas to the DFT-MD data. The virial coefficients finally used are given in Tab.~\ref{tab:virial}.

For temperatures below 600~K we used the classical results for $B_2(T)-B_5(T)$ from \cite{Shaul2012} derived from state-of-the-art Mayer-sampling Monte Carlo calculations.
For temperatures between 600~K $\leq$ T $\leq$ 10~kK we found smoothest connections using the \textit{ab initio} virial coefficients $B_2(T)$ and $B_3(T)$ of \cite{Bich2007}. 
They derived their results from a high-quality 18-parameter \textit{ab initio} interaction potential for helium, see \cite{Hellmann2007} for details.

 \begin{deluxetable}{rcccc}
 \tablecolumns{2}
 \tablewidth{0pc}
 \tablecaption{Virial coefficients B$_i$ taken from \cite{Shaul2012} by interpolation of their given data (labeled with a) and from \cite{Bich2007} where no interpolation is needed (labeled with b) which we used for the virial part of our helium EOS.}
 \tablehead{ \colhead{T [K]} & \colhead{B$_2$ $\left[\frac{\mathrm{cm}^3}{\mathrm{mol}}\right]$}& \colhead{B$_3$ $\left[\frac{\mathrm{cm}^6}{\mathrm{mol}^2}\right]$} 
 & \colhead{B$_4$ $\left[\frac{\mathrm{cm}^9}{\mathrm{mol}^3}\right]$} & \colhead{B$_5$ $\left[\frac{\mathrm{cm}^{12}}{\mathrm{mol}^4}\right]$}  }
 \startdata
 60$^a$ & 7.46483 & 161.676 & 1553.11 & 15478.4\\
 100$^a$ & 10.4982 & 147.007 & 1322.62 &9857.49 \\
 200$^a$ & 11.6929 & 121.309 & 882.001 & 4458.25\\
 300$^a$ & 11.5403 & 104.351 & 650.567 &2547.36 \\
 600$^b$ & 10.651 & 78.73 & \nodata & \nodata\\
 1000$^b$ & 9.5497 & 59.44 & \nodata & \nodata\\
 2000$^b$ & 7.9556 & 38.60 & \nodata & \nodata\\
 3000$^b$ & 7.0330 & 29.16 & \nodata & \nodata\\
 6000$^b$ & 5.5459 & 17.14 & \nodata & \nodata\\
 10000$^b$ & 4.5542 & 11.05 & \nodata & \nodata
 \enddata
 \label{tab:virial}
 \end{deluxetable}

We calculated the ideal parts $P^{\mathrm{id}}$ and $u^{\mathrm{id}}$ of the virial EOS from the free energy of an ideal plasma model as described in \cite{Foerster1992}:
\begin{equation}
 F = Nk_BT\left[\sum \limits_{z=0}^2 \alpha_z\mathrm{ln}\left(\frac{n_z\lambda_z^3}{\sigma_z}\right) -1\right].
 \label{eq:id}
\end{equation}
The index z represents the z-fold charged atoms with a relative fraction of $\alpha_z=n_z/n$ with respect to the total atom density $n=N/V=n_0+n_1+n_2$.
The thermal wavelength of the atoms is given by $\lambda_z=h(2\pi m_zk_BT)^{-1/2}$ and the internal partition functions $\sigma_z$ were calculated using the Planck-Larkin convention~\citep{Planck1924,Larkin1960}:
\begin{eqnarray}
 \sigma_z  &=& \exp\left( -\sum \limits_{z^\prime=0}^z \frac{I_{z^\prime}}{k_BT} \right) \\ 
 && \times\sum_m g_{z,m}\left[ \exp\left(\frac{-E_{z,m}}{k_BT}\right)-1+\frac{E_{z,m}}{k_BT} \right] \;,\nonumber\\
 && z=0,1 \quad,\quad \sigma_2 = \exp\left( -\sum \limits_{z^\prime=0}^2 \frac{I_{z^\prime}}{k_BT} \right)  \;.
\end{eqnarray}
The ionization energies $I_z$, the energy levels $E_{z,m}$, and the corresponding statistical weights $g_{z,m}$ are taken from the 
NIST database \citep{NIST2013}. Note that here is one difference to the SCvH-He EOS, see below, where only the ionization energies $I_z$ are 
considered in the internal contribution to the free energy while the $E_{z,m}$ terms are neglected. 

The relative fractions of the atoms $\alpha_z$ are determined by coupled Saha equations
\begin{equation}
 \mu_{0} = \mu_{1} + \mu_{e^-} \quad,\quad \mu_{1} = \mu_{2} + \mu_{e^-} \;,
 \label{eq:Saha}
\end{equation}
with the boundary condition of charge neutrality of the system: $n_e=n_{1}+2n_{2}$. 
$P^{\mathrm{id}}$ and $u^{\mathrm{id}}$ are then obtained by differentiation of the specific free energy ($f=F/m$):
\begin{equation}
 P = \rho^2\left(\frac{\partial f}{\partial \rho} \right)_T \;,\; 
 u = f-T\left(\frac{\partial f}{\partial T} \right)_\rho \;.
\end{equation}
 
For temperatures above 10~kK helium starts to ionize. The additional ideal contributions to the pressure and the internal energy from the unbound electrons lead to a non-horizontal behavior of the ideal gas isotherms, see Figs.~\ref{fig:He10MKp} and~\ref{fig:He10MKu}. 

It turns out that the helium EOS of \cite{Saumon1995} (SCvH-He) is the most appropriate one above 10~kK for our purposes: describing the partial and fully ionized helium with interaction and correlation effects and a smooth transition into the DFT-MD data. The full description of their EOS can be found in their paper, while we only focus on the main points for a better understanding of the composition of our EOS table. The free energy of their model can be written as
 \begin{equation}
  F(n_0,n_1,n_2,n_e,V,T) = F_{\mathrm{id}} + F_{\mathrm{conf}} + F_{\mathrm{DH}} + F_{\mathrm{int}} \;.
 \end{equation}
$F_{\mathrm{id}} + F_{\mathrm{int}}$, the ideal and internal part, is in principle the same expression as Eq.~(\ref{eq:id}), but with a simpler approximation for the internal partition function -- it only takes into account the ionization energies $I_z$ and treats electrons on the level of the Fermi integral in order to include degeneracy effects. The contribution $F_{\mathrm{conf}}$ is due to the interaction of the helium atoms and calculated via fluid perturbation theory. The interaction of charged particles is treated using the Debye-H\"uckel approximation $F_{\mathrm{DH}}$. The fully ionized plasma for densities above 3~g/cm$^3$ is described by a screened one-component plasma model \citep{Chabrier1990}.

We extended the SCvH-He isotherms down to $10^{-10}$~g/cm$^3$ (smallest density of the virial EOS) using the ideal plasma model that connects almost perfectly to the SCvH-He data. That is why the respective red area in Fig.~\ref{fig:HeArea} is labeled as "SCvH-EOS+ideal plasma limit". As for hydrogen the connection of the partially ionized plasma to the DFT-MD data becomes difficult when significant ionization takes place in the transition region. For helium this is the case between 60~kK and 300~kK where we applied spline interpolation for generating a smooth transition between the SCvH-He and the DFT-MD data, see Fig.~\ref{fig:HeArea}.
 
\begin{figure}
 \plotone{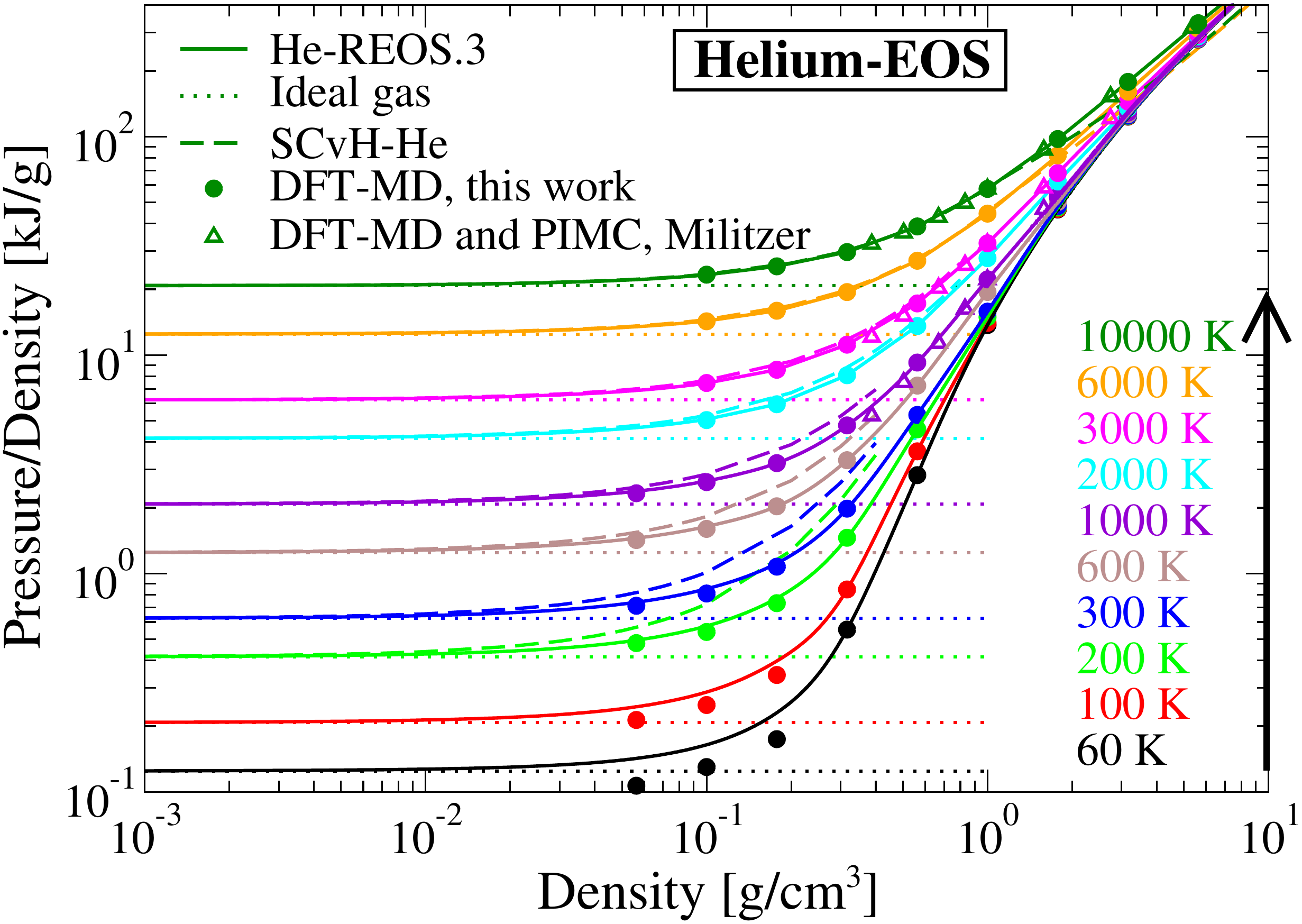}
 \caption{\label{fig:He10kKp}He-REOS.3 isotherms for 60~K to 10~kK. Displayed is the ratio of pressure and density over the density. Solid lines are the He-REOS.3, dotted lines represent the ideal gas limit, dashed lines show the SCvH-He EOS while our DFT-MD data are labeled with circles; DFT-MD and the PIMC results from Militzer (open triangles) are for comparison.}
\end{figure}
 
\begin{figure}
 \plotone{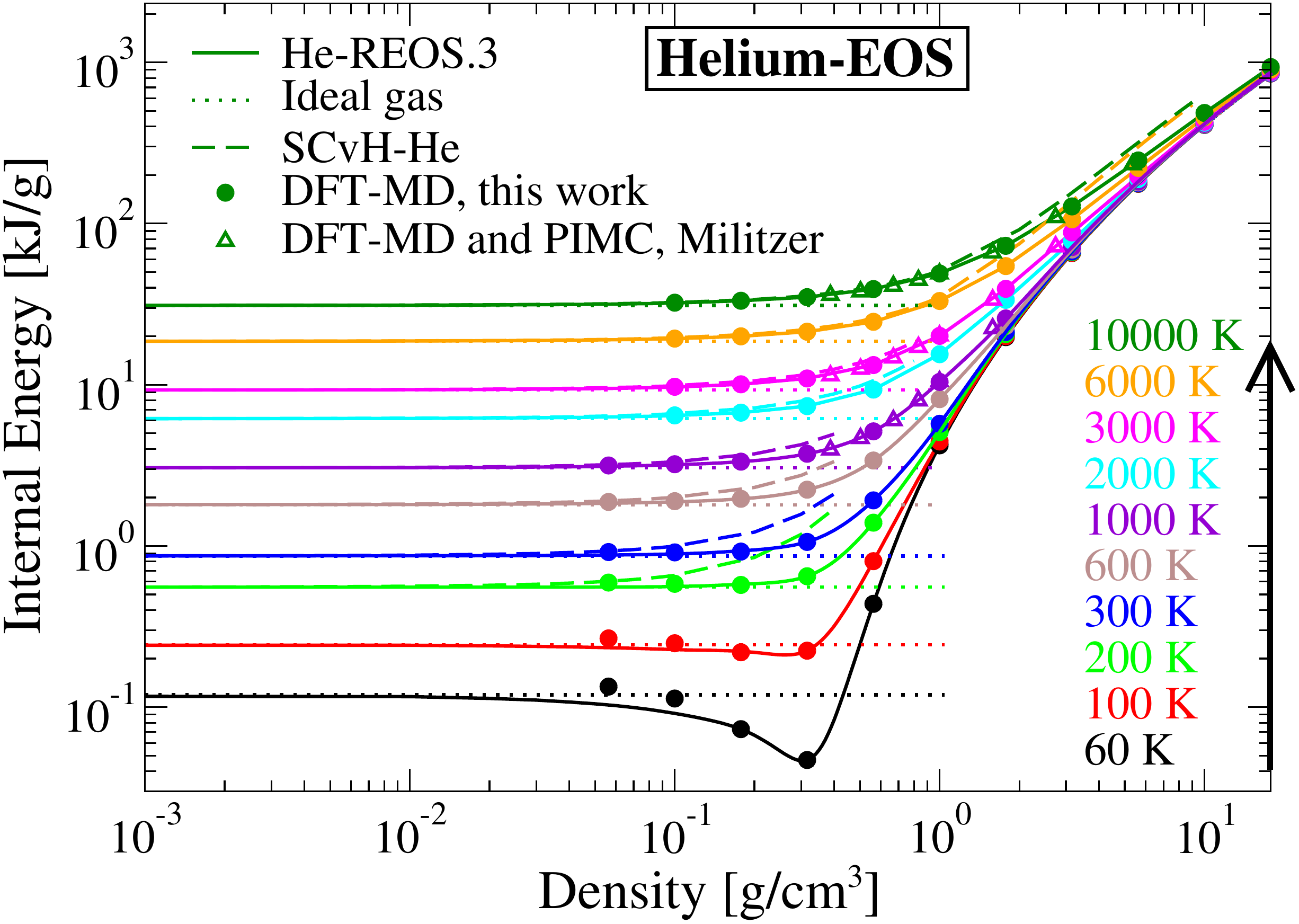}
 \caption{\label{fig:He10kKu}He-REOS.3 isotherms for 60~K to 10~kK. Displayed is the internal energy versus the density. The line styles are associated with the same labels as in Fig.~\ref{fig:He10kKp}.}
\end{figure}

\begin{figure}
 \plotone{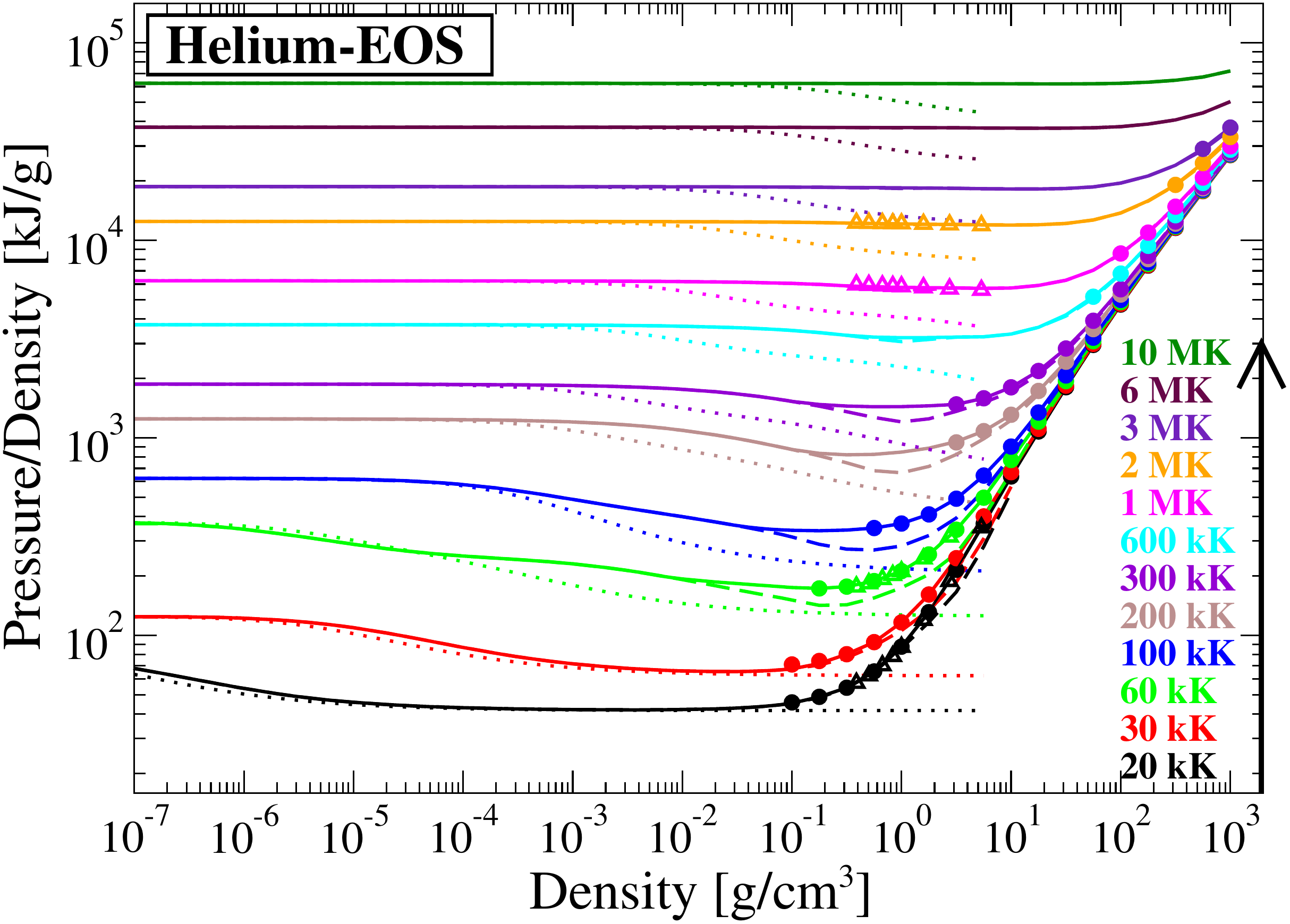}
 \caption{\label{fig:He10MKp}He-REOS.3 isotherms for 20~kK to 10~MK. Displayed is the ratio of pressure and density versus the density. The line styles are associated with the same labels as in Fig.~\ref{fig:He10kKp}.}
\end{figure}
 
\begin{figure}
 \plotone{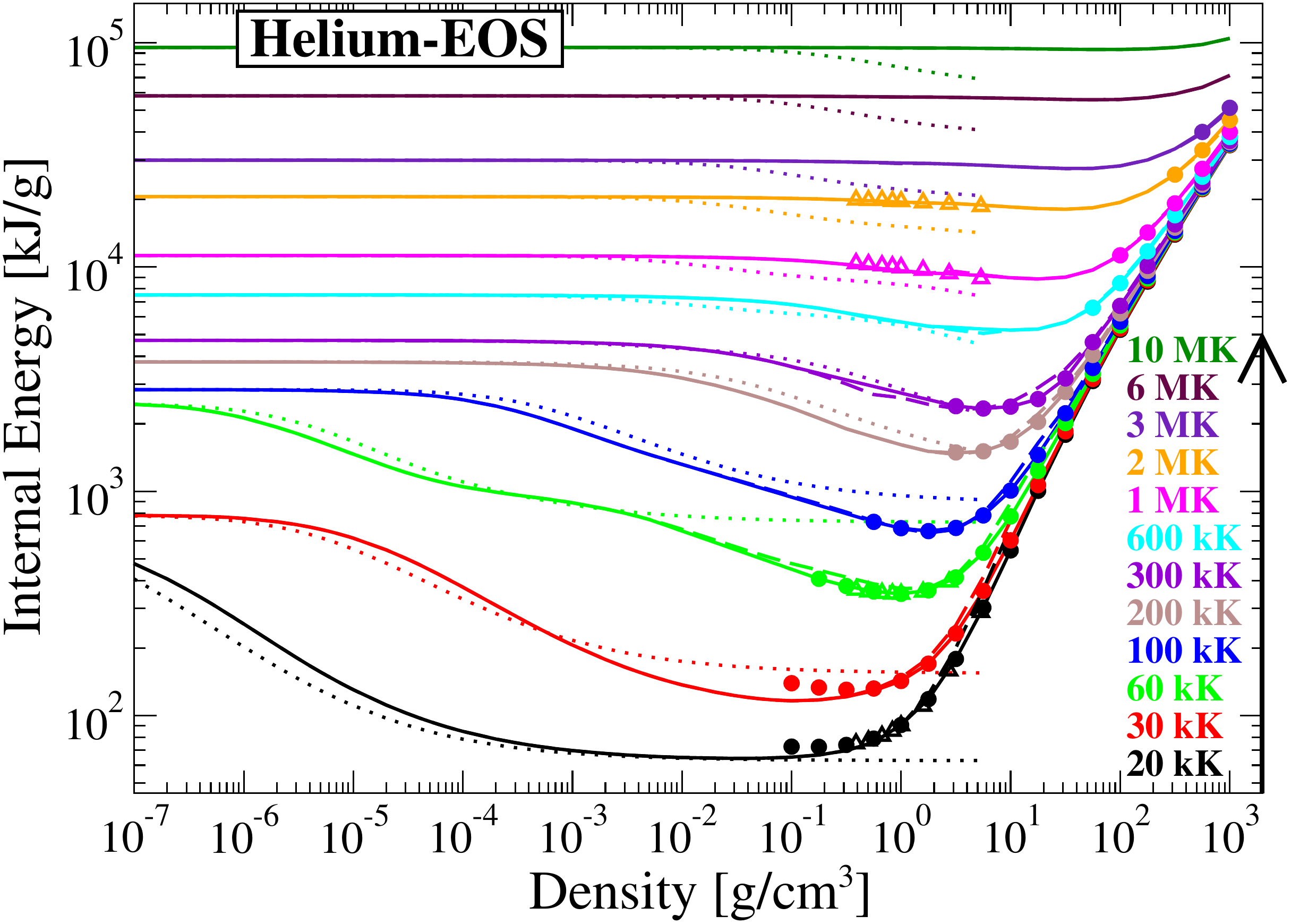}
 \caption{\label{fig:He10MKu}He-REOS.3 isotherms for 20~kK to 10~MK. Displayed is the internal energy versus the density. The line styles are associated with the same labels as in Fig.~\ref{fig:He10kKp}.}
\end{figure}

\subsubsection{EOS table composition}

Having presented the constitutive parts of our EOS tables in the previous section, we now explain how we constructed the final EOS tables. The requirements of an accurate description of the underlying physics, of preserving thermodynamic consistency, and obtaining smooth transitions between the various models could always be met within a reasonable compromise.
The isotherms of each model are shown in Figs.~\ref{fig:He10kKp}-\ref{fig:He10MKu}. The pressure isotherms are again displayed in the $P(\rho,T)/\rho$ representation that reduces the ideal contribution to a constant offset for each isotherm and stresses the non-ideal contributions to the EOS.

As mentioned above, for temperatures below 20~kK the transition from the ideal gas to the DFT-MD data has been performed via the virial EOS, see Figs.~\ref{fig:He10kKp} and~\ref{fig:He10kKu} where the contribution of the virial coefficients are in principle those parts of the isotherms that differ from the ideal gas (dotted) horizontal lines below 0.1~g/cm$^3$ above 300~K. There, the virial expansion connects right into the lowest DFT-MD data. Taking into account pressure and energy isotherms below 300~K it is reasonable to use the virial EOS up to 0.3~g/cm$^3$, because DFT-MD simulations are challenging there. For example at 60~K the virial expansion shows the right negative slope in internal energy, indicating the transition to the solid, that is covered by the DFT-MD data above 0.3~g/cm$^3$. However, the respective DFT-MD pressures are partly below the ideal gas that is not indicated by the virial expansion. That is why the virial expansion is favored in this regime.

As shown in Figs.~\ref{fig:He10kKp} and~\ref{fig:He10kKu}, we find a very good agreement with the \textit{ab initio} data from \cite{Militzer2009} (open triangles) except small deviations in pressure for the two lowest densities below 10~kK. EOS data from the SCvH-He model (dashed lines) overestimate the pressure significantly below 3000~K and densities around 0.1~g/cm$^3$ and overestimate the internal energy for all isotherms up to 10~kK at the same densities. Note that the SCvH isotherms merge into the corresponding DFT-MD data for temperatures above 10~kK at the transition densities. That is why we prefer this model to extend our EOS table into the partial and fully ionized regime of the density-temperature plane, see Figs.~\ref{fig:He10MKp} and \ref{fig:He10MKu} to which we focus our discussion now.

The ideal gas model includes by definition no interaction effects and hence no pressure ionization at higher densities. This leads to a false prediction of an atomic system for increasing density caused by the higher recombination probabilities within the Saha equations~(\ref{eq:Saha}), leading to a drop of the pressure and energy isotherms (dotted curves). However, the ideal plasma model is exact in the low-density limit and serves, therefore, as the extension of the ideal SCvH-He data down to 10$^{-10}$~g/cm$^3$.

The most difficult connection between the SCvH-He and DFT-MD data occured for temperatures between 60~kK and 300~kK, see Fig.~\ref{fig:HeArea}. This is caused by the 
ionization of helium and the effects of recombination and pressure ionization for higher densities. While we used the SCvH-He data at the 20~kK and 30~kK energy 
isotherms up to 0.5~g/cm$^3$, we had to apply interpolations, in particular for the pressure isotherms from 60~kK on.
In detail, the following data points of the He-REOS.3 table are generated using natural cubic spline interpolations: 0.01-0.1~g/cm$^3$ at 60~kK, 0.05-0.3~g/cm$^3$ at 100~kK, 0.1-3~g/cm$^3$ at 200~kK, and 0.1-3~g/cm$^3$ at 300~kK.
This interpolated data can be identified in Figs.~\ref{fig:He10MKp} and \ref{fig:He10MKu} where the solid He-REOS.3 curves deviate from the SCvH-He data (dashed lines). These data underestimate the pressures systematically in the relevant region predicted from our simulations (circles) and those of \cite{Militzer2009} (open triangles), which coincide.

As shown in Figs.~\ref{fig:He10MKp} and \ref{fig:He10MKu} the transition from the SCvH-He model to the DFT-MD data for temperatures above 300~kK proceeds very smoothly, while all \textit{ab initio} data (circles and open triangles) are located on the final He-REOS.3 isotherms (solid lines). Thus, no interpolations were performed here. The 6~MK and 10~MK isotherms are pure SCvH-He data which proceed almost horizontally in the representations and, thus, are nearly perfect ideal gas isotherms.

\subsection{Normalization of the internal energy}\label{subsec:Normalization}
The zero point $u_0$ of the specific internal energy $u(\rho,T)$ can be chosen arbitrarily. We decided to fix our EOS tables to the limiting case of the perfect ionized ideal plasma at very high temperatures and very low densities. One can simply show that this limiting case is material dependent and obeys the following equation:
\begin{equation}
 u=\frac{3}{2}\frac{Nk_BT}{m}=\frac{3}{2}\frac{k_BT}{\rho}(n_e+n_i) =\frac{3}{2}\frac{(Z+1)RT}{M_{at}}.
\end{equation}
It is $N$ the total particle number of the system, $k_B$ the Boltzmann constant, $n_e$ and $n_i$ the electron- and ion-densities of the system, $Z$ the atomic number and $M_{at}$ the molar mass of the atomic system.
For hydrogen with $Z = 1$ and $M_{at} = 1$~g/mol we fixed the zero point to $u_0(10\mathrm{~MK}) = 249.433$~MJ/g. For helium with $Z = 2$ and $M_{at} = 4$~g/mol we obtained $u_0(10\mathrm{~MK}) = 93.538$~MJ/g. Therefore, at the lowest densities at 10~MK in the HREOS.3 and He-REOS.3 tables the specific internal energy has the respective $u_0$ values, see Tabs.~\ref{tab:H-REOS} and~\ref{tab:He-REOS}.
\subsection{Thermodynamic consistency}\label{subsec:TdynCons}

A sensible test for the accuracy of our REOS.3 tables is checking thermodynamic consistency that is given by Eq.~(\ref{eq:fundamental}). To get a relative deviation with respect to the pressure of the system we divide Eq.~(\ref{eq:fundamental}) by $P$ and subtract unity which leads to
\begin{equation}
 \frac{T}{P}\left(\frac{\partial P}{\partial T} \right)_\rho + 
 \frac{\rho^2}{P}\left(\frac{\partial u}{\partial \rho} \right)_T -1 = \Delta \:.
 \label{eq:Tdynconsi}
\end{equation}
For instance, $\Delta=0$ denotes perfect thermodynamic consistency while $\Delta=0.1$ implies a violation of $10\%$. The derivatives that are needed for the evaluation of Eq.~(\ref{eq:Tdynconsi}) are obtained from a more resolved EOS grid generated via cubic spline interpolation. In detail we used natural cubic splines for interpolations along isotherms because the spacing between the original density grid is small. Unlike the original temperature grid that have certain jumps, e.g. for helium with 6~kK, 10~kK, 20~kK. Here we applied cubic Akima splines that do not tend to overshoot interpolating this rougher grid.

\begin{figure}
 \plotone{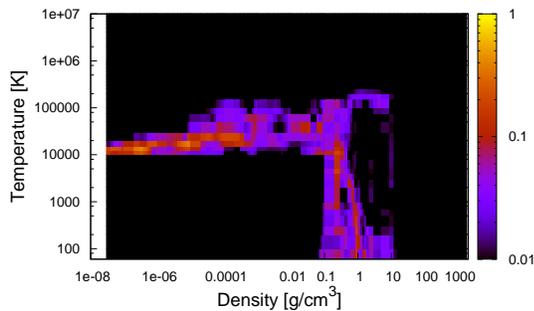}
 \caption{\label{fig:HConsi}Thermodynamic consistency of the entire hydrogen EOS. Black areas indicate that the relative deviation of $P(\rho,T)$ and $u(\rho,T)$ from the fundamental relation (\ref{eq:fundamental}) is below $1\%$.}
 \end{figure}

\begin{figure}
 \plotone{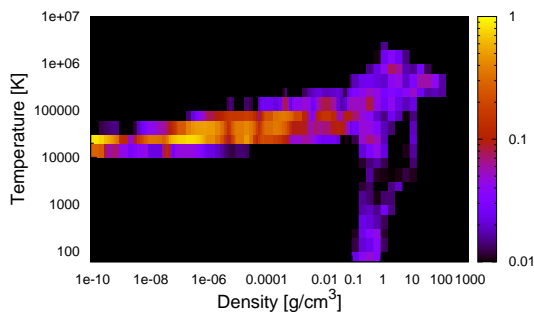}
 \caption{\label{fig:HeConsi}Thermodynamic consistency of the entire helium EOS. Black areas indicate that the relative deviation of $P(\rho,T)$ 
 and $u(\rho,T)$ from the fundamental relation (\ref{eq:fundamental}) is below $1\%$.
 }
\end{figure}

The check of thermodynamic consistency for the H-REOS.3 and He-REOS.3 is shown in Fig.~\ref{fig:HConsi} and Fig.~\ref{fig:HeConsi}, respectively. The color code represents areas where thermodynamic consistency is ensured within a certain accuracy. For instance, a performance with errors $\leq1\%$ is given in black, $\sim10\%$ in red, and $\sim100\%$ in yellow.
As it was expected, most inconsistencies occur at the boundary between two EOS models at 10~kK where ionization occurs and in the interpolation regimes. Note that the SCvH model contains thermodynamic inconsistencies within their interpolation areas as well. Interestingly, we obtain red areas in particular in the transition region from 10~kK to 20~kK at low densities for both EOS tables. This can be explained directly by our criterion. This transition region is within or near the ideal gas regime but at the points where ionizations becomes increasingly important, leading to a slope in the pressure and energy isotherms, see Figs.~\ref{fig:He10kKp}-\ref{fig:He10MKu}.
This slope has a strong influence on the isochores with respect to the interpolation scheme that takes into account the two neighboring points for each data point. For instance, for helium we have the 6~kK, 10~kK, 20~kK, 30~kK and 60~kK isotherms, which is a rather big spacing for interpolation and obtaining the quite sensitive derivative $(\partial P/\partial T)_\rho$, see Eq.~(\ref{eq:Tdynconsi}). This term has to compensate the $(\partial u/\partial \rho)_T$ term, which is zero within the ideal gas model because its caloric EOS is independent of the density.
Hence, the red areas in the ideal gas regime should be considered as intrinsic effects of our consistency criterion, producing ostensible inconsistencies. However, they are not important for the calculation of isentropes for Giant Planets and Brown Dwarfs.
Remarkably, these thermodynamic inconsistencies are absent at the transition region from the SCvH model to the EOS of Chabrier-Potekhin in the Hydrogen EOS, see Figs.~\ref{fig:HConsi} and~\ref{fig:HArea}.
Since the hydrogen is fully ionized there, no significant changes in the slope of the neighboring isotherms occur.

Finally, we conclude that in those parts of the H-REOS.3 and He-REOS.3 which are relevant for calculating isentropes for GPs and BDs (see Figs.~\ref{fig:HArea} and~\ref{fig:HeArea}), thermodynamic consistency is ensured to be better than $5\%$, in most cases better than $1\%$.

\subsection{Comparison with high-pressure experiments}\label{subsec:Experiments}
High-pressure experiments using static diamond anvil cells or dynamic shock compression techniques are a crucial test for each EOS. Since an extensive comparison of 
our DFT-MD data for hydrogen with a variety of experiments is given in \cite{Becker2013} we will focus here on the helium EOS. In contrast to hydrogen, only a few 
high-pressure experiments for helium are published.
The cold curve at 300~K has been investigated by several static anvil compression experiments \citep{Lallemand1977, Mills1980, Loubeyre1993, Mao1988, Eggert2008}.
We find a very good agreement with our He-REOS.3 and these experimental results. In particular our data coincide with the experiments below 0.3~g/cm$^3$. At higher densities our pressures are systematically higher than in the experiments with a maximum deviation from the Loubeyre data in the solid of 10\% at 1.5 g/ccm. 

\begin{figure}
 \plotone{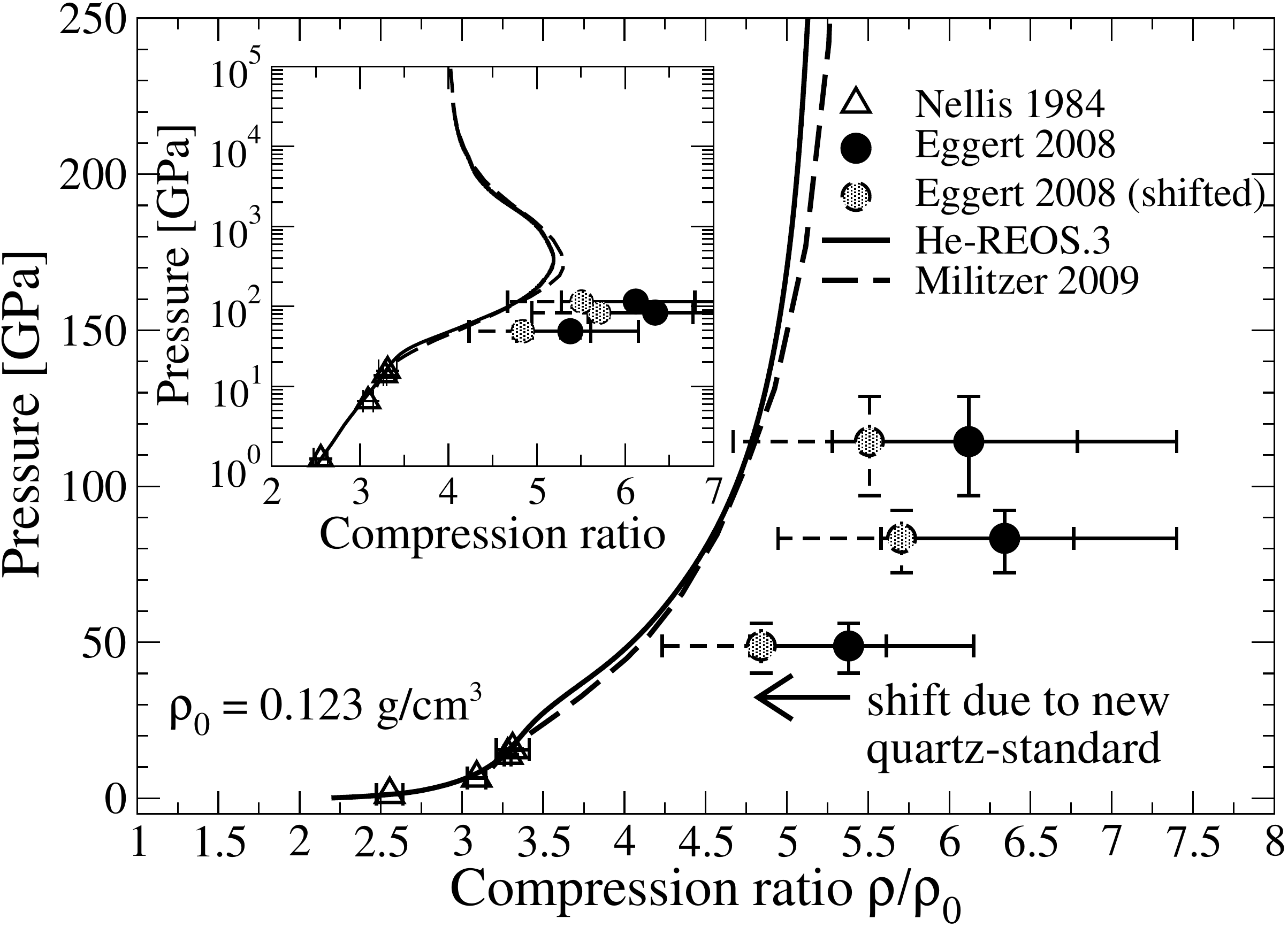}
 \caption{\label{fig:HugoPres}The principal Hugoniot of helium with an initial density of $\rho_0=0.123$~g/cm$^3$ at 4~K: Experimental results from \cite{Nellis1984} (triangles) and from \cite{Eggert2008} (circles) are shown together with predictions from our DFT-MD data (solid line) and the DFT-MD and PIMC data of \cite{Militzer2009} (dashed line). The shift of the Eggert data are explained in the text. }
\end{figure}

To test our EOS at higher temperatures more relevant to the BD regime, we compare to the principal Hugoniot experiments of \cite{Nellis1984, Eggert2008} and 
\cite{Celliers2010}. Of particular interest are the gas gun shots from \cite{Nellis1984} that yielded a pressure of 16~GPa at 0.41~g/cm$^3$ and $\sim12000$~K. This is an EOS point very close to the adiabat of Gliese-229b, see Sec.~\ref{sec:IntBrownie}.
 The experimental data for the principal Hugoniot together with theoretical predictions from our He-REOS.3 (solid) and the DFT-MD and PIMC results of \cite{Militzer2009} 
 (dashed) are shown in Fig.~\ref{fig:HugoPres}. Note that the data of \cite{Eggert2008} (circles) rely on the EOS of the reference material (quartz) used in the experiment.
 This EOS has been recalibrated by \cite{Knudson2009}, leading to a density correction of the original helium results. \cite{Celliers2010} estimated this 
 correction to be $10\%$, leading to the shift from the filled to the shaded circles in Fig.~\ref{fig:HugoPres}.
 
 The \textit{ab initio} predictions agree very nicely for lower and high pressures but our He-REOS.3 is stiffer at intermediate pressures which yields a smaller maximum 
 compression, see inset of Fig.~\ref{fig:HugoPres}. Both theoretical curves coincide with the experimental results of \cite{Nellis1984} but only with the shifted highest 
 compression of \cite{Eggert2008}. This issue might be solved with a detailed reanalysis of their data based on the new quartz standard.
 We point out that the relevant experimental data for shallow depths of BDs \citep{Nellis1984} are matched by our He-REOS.3.

\subsection{Linear and real mixtures}\label{subsec:LMvsRM}

The major constituents of GPs and BDs are hydrogen and helium and a small fraction of heavier elements. Hence, interior models are based on a mixture EOS of them. 
As we provide two seperate EOS tables for hydrogen and helium, we apply the additive volume rule to obtain a linear mixture H-He EOS (LM-EOS) with the mass fraction of helium (Y) and hydrogen ($X=1-Y$):
\begin{eqnarray}
 \frac{1}{\rho_{mix} (P,T)} & = & \frac{X}{\rho_{H} (P,T)} + \frac{Y}{\rho_{He} (P,T)} \;,\label{eq:LM1}\\
 u_{mix}(P,T) & = & Xu_H(P,T) + Yu_{He}(P,T) \;.\label{eq:LM2}
\end{eqnarray}
Taking into account a representative EOS for the heavier elements (Z), such as water for GPs or a scaled helium EOS as done within the BD calculations 
in Sec.~\ref{sec:IntBrownie}, $Z/\rho_{Z}(P,T)$ has to be added to the right hand side of Eq.~(\ref{eq:LM1}) and $Zu_Z(P,T)$ to Eq.~(\ref{eq:LM2}). 
The LM-EOS can then be applied for arbitrary compositions $X/Y/Z$. 

\begin{figure}
 \plotone{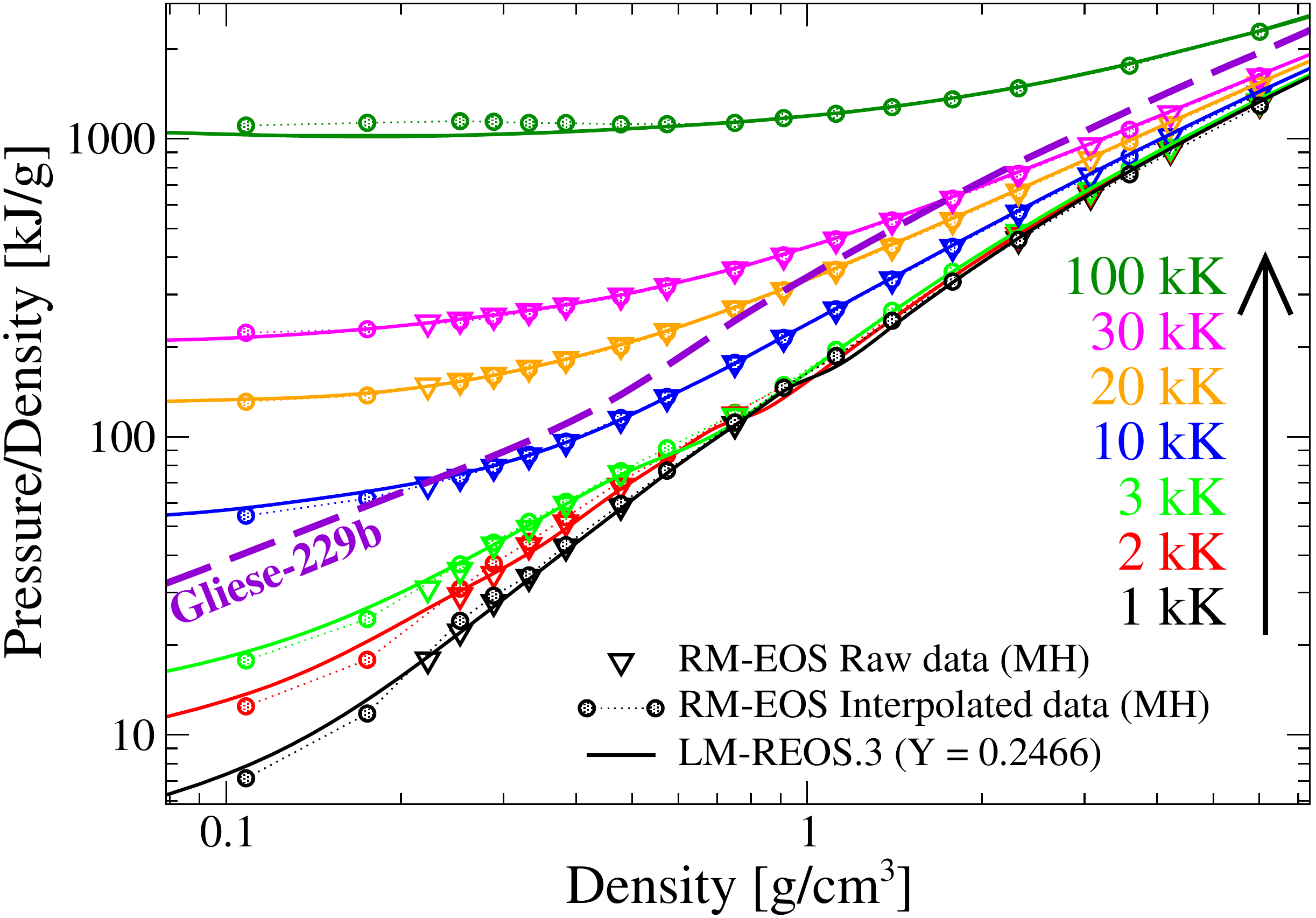}
 \caption{\label{fig:MilipDiv}Linear mixture (LM: solid lines) and real mixture (RM) of hydrogen and helium with $Y=0.2466$. Open triangles represent the DFT-MD data of \cite{Militzer2013b} (MH) and the circles their interpolated EOS data. The dashed violet curve is the adiabat of Gliese-229b, see Sec.~\ref{sec:MRR}.}
\end{figure}

\begin{figure}
 \plotone{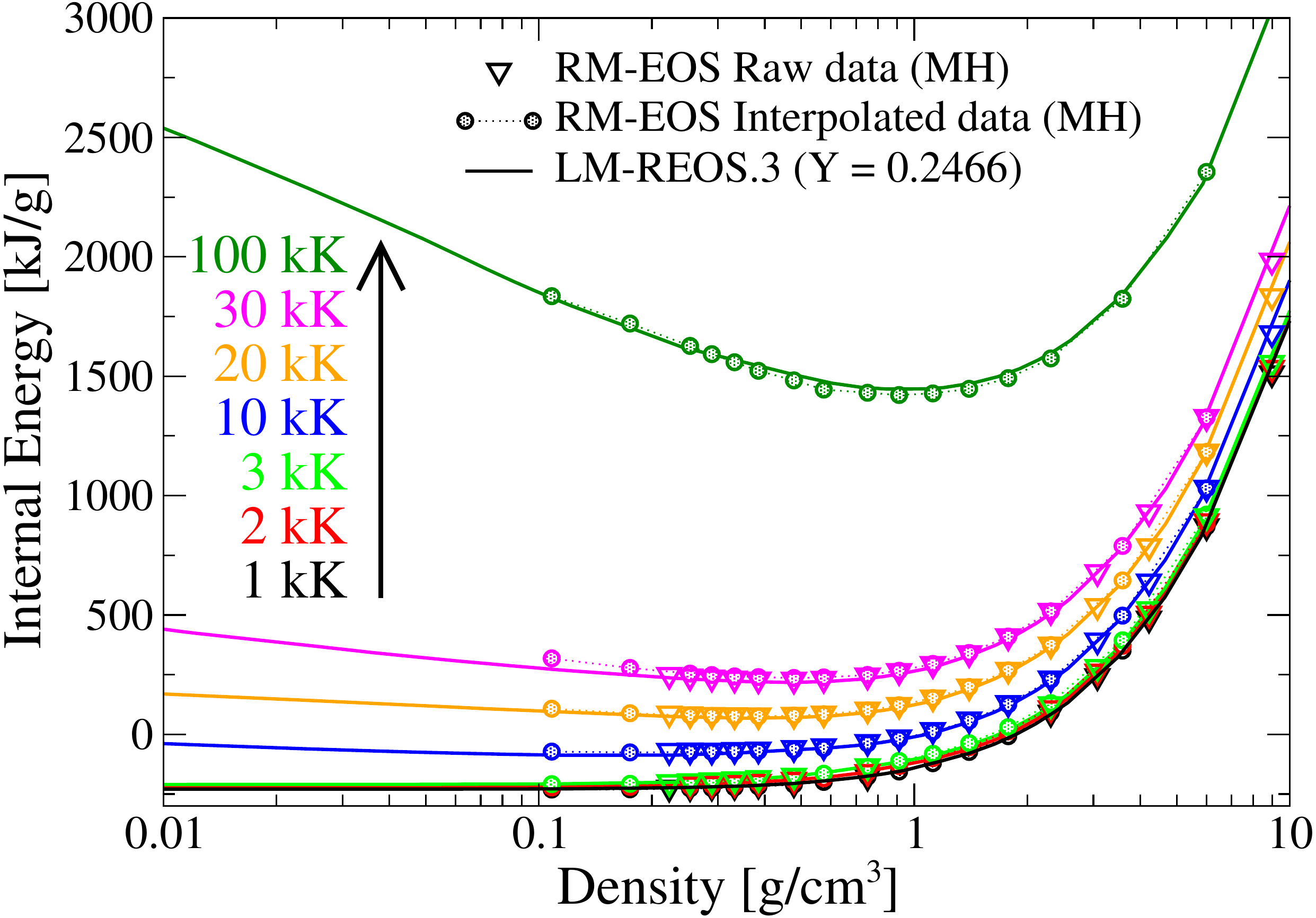}
 \caption{\label{fig:MiliuLin} Energy isotherms of the linear (LM) and real hydrogen-helium mixture. The line styles are associated with the same labels as in Fig.~\ref{fig:MilipDiv}.}
\end{figure}

The alternative approach is to perform \textit{ab initio} simulations for a mixture of hydrogen and helium (and perhaps heavier elements) with the desired fractions to obtain the EOS for given values of $X/Y/Z$, see \cite{Lorenzen2009}. This leads to the real mixture EOS (RM-EOS) which includes non-linear mixing effect in a genuine way.
In a recent paper \cite{Militzer2013b} provide EOS data for a real mixture with a helium fraction of $Y=0.2466$.
Pressure and energy isotherms of their original DFT-MD data (open triangles) and their interpolated data (circles) are shown in Figs.~\ref{fig:MilipDiv} and~\ref{fig:MiliuLin} together with our LM isotherms for this $Y$ value. Note that the 100~kK RM isotherm and all RM data below 0.2~g/cm$^3$ are extrapolated. Furthermore, we do not see a first-order phase transition in the original RM data for hydrogen around 1~g/cm$^3$ below 2000~K as has been predicted recently~\citep{Lorenzen2010,Morales2010a}.

We find a remarkably good agreement of our LM-REOS.3 with the original RM data, especially in the region above 10~kK that is relevant for the interior of BDs. As an example we show the adiabat of Gliese-229b (dashed violet curve in Fig.~\ref{fig:MilipDiv}). We conclude that non-linear mixing effect are not relevant for $\rho-T$ conditions that occur in BDs but may be important for GPs as stated in 
\cite{Militzer2013b}, see next section. Therefore, we apply our LM-REOS.3 for modeling the interior of BDs.

Another important phenomenon that occurs in real H-He mixtures is the demixing of both components at high pressures. This issue has been discussed for decades~\citep{Stevenson1977a, Lorenzen2011, Morales2013b}. It affects the interiors of rather cold objects such as Saturn and presumably Jupiter~\citep{Stevenson1977b, Fortney2004a}. Caused by demixing, helium droplets can form and sink into the planet, thereby releasing gravitational energy as an additional energy source. This process can explain the luminosity problem of Saturn~\citep{Fortney2003}, where homogeneous evolution models predict an age of $\sim 2.5$~Gyr in contradiction to the age of the solar system of 4.56~Gyr. This problem could be solved alternatively assuming a layered convection for the interior structure of Saturn as proposed by~\cite{Leconte2013}.

However, a demixing effect is not included in a linear mixture of our EOS data. Therefore we predict a thermodynamic and mass range for homogeneous objects consisting of hydrogen and helium with a solar helium fraction where demixing will occur in their interiors. The results are based on the demixing regions calculated from \textit{ab initio} simulations by~\cite{Lorenzen2011} and~\cite{Morales2013b} who predict the phase separation above 1~Mbar for temperatures below 8000~K. These conditions are fulfilled in the interiors of objects with a mass $>0.125$~M$_\mathrm{Jup}$ and a temperature at the 100~bar level of $T(100~\mathrm{bar})\leq700$~K. For comparison, the temperatures at 100~bar for the massive Giant Planet KOI-889b and the Brown Dwarfs Corot-3b, Gliese-229b and Corot-15b which we will investigate in Sec.~\ref{sec:IntBrownie} are 1100~K and 2500~K, well above the demixing region.

\section{Jupiter models using the LM-REOS.3 data}\label{sec:Jupi}

As a first application and benchmark of our hydrogen and helium EOS data we calculate interior models for Jupiter based on a three-layer structure similar to previous studies~\citep{Nettelmann2012} where details of the interior models can be found. We only mention the key assumptions here. The planet is assumed to consist of a rocky core surrounded by two adiabatic fluid layers which differ in composition; the variable layer boundary is located at the transition pressure $P_{12}$. The adiabatic boundary condition is fixed at the 1-bar level: $T(1~\textrm{bar})=170$~K. The interior models have to reproduce observational constraints: the Jovian mass and its radius, the 1-bar temperature, the atmospheric mass abundance of helium ($Y_1=0.238$), the solar mean helium abundance~$\overline{Y}$, the angular velocity, and the three lowest-order gravitational moments $J_2$, $J_4$, $J_6$.

The main finding of \cite{Nettelmann2012} was higher possible values for the heavy element abundance in Jupiter's outer envelope ($Z_1$) and atmosphere using H-REOS.2 instead of H-REOS.1. In particular, the maximum possible $Z_1$ value was found to be 2.7 times the solar value of 1.49~\citep{Lodders2003}, in good agreement with the in-situ measured noble gas abundance of $\sim2\times$~solar, but still lower than the measured carbon abundance.

\begin{figure}
 \plotone{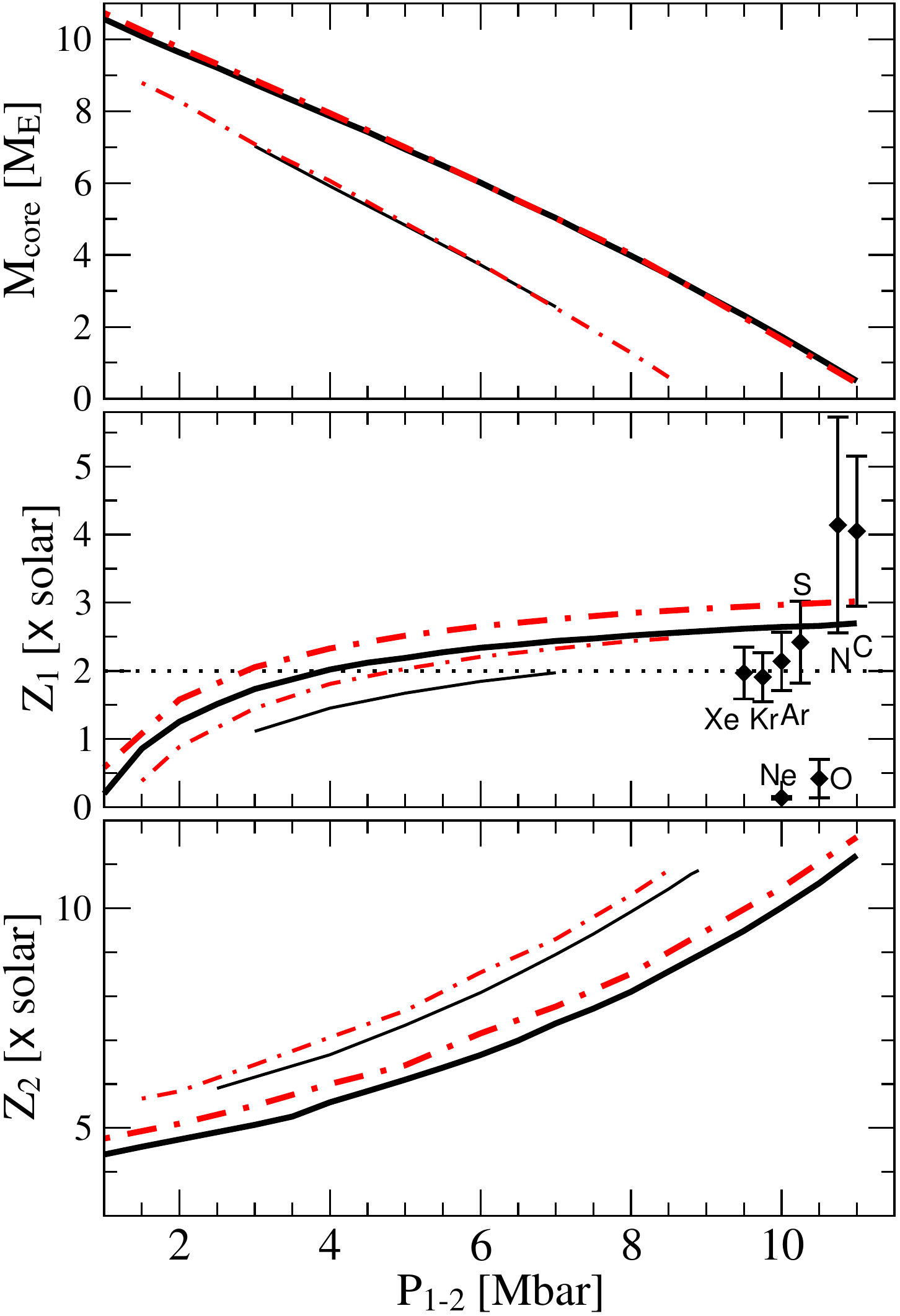}
 \caption{\label{fig:Jupi}Core mass and metallicities in the inner and outer layer plotted over the transition pressure. Black lines are taken 
 from~\cite{Nettelmann2012} as reference. Results using the new REOS.3 data are shown in red. 
 The thick (thin) curves are calculated with the theory of figures to third (fourth) order. 
}
\end{figure}

Using our new H and He equations of state, we find a slight but important enhancement of the possible maximum $Z_1$ value up to $3\times$~solar, compare the red lines (new results) to the black lines \citep{Nettelmann2012}
 in the middle panel of Fig.~\ref{fig:Jupi}. An atmospheric metallicity of $3\times$~solar is consistent with the $3-5\times$~solar enrichment of carbon, which is supposed 
 to be one of the most abundant elements in Jupiter and thus may serve as a representative for Jupiter's atmospheric metallicity as long as oxygen abundance remains unknown.
 We note that the core mass of the models remains unchanged using the improved equations of state.

The increase of the metallicities can be explained as follows. The gravitational moments that are fitted by adjusting $Z_1$ and $Z_2$ are most sensitive around a few Mbars ($J_2$) or below 1 Mbar ($J_4$).
Since the raw data for the conditions in the outer envelope of both hydrogen EOS (H-REOS.2 and H-REOS.3) are nearly the same, the increase in $Z_1$ can only be attributed to the new helium EOS.
Indeed, our new helium EOS is less compressible than the former one, so that more heavy elements must be added to obtain the same mass density as constrained by the gravitational moments. 

It was mentioned in the previous section that non-linear mixing effects which are absent in our LM-REOS.3 might play a role for Giant Planets like Jupiter as argued in~\cite{Militzer2013b}. However, a fully converged Jupiter model fulfilling all observational constraints that is simultaneously based on a real-mixture isentrope is still unavailable. Instead we have compared the Jupiter-like isentrope provided in~\cite{Militzer2013b} with a LM-REOS.3 isentrope starting from the same initial condition ($P(3770\mathrm{~K}) = 10.44$~GPa). In the pressure range between 0.4-4~Mbar, where the gravitational moments are most sensitive~\citep{Guillot2005}, we found a maximum deviation of 4\% which increases systematically up to 9\% at the core-mantle boundary at $\sim40$~Mbar. Therefore, we provide interior models based on a linear mixture that satisfy all observational constraints so that the discussion about the importance of non-linear mixing effect remains open.

\section{Mass-radius relations and interior models for brown dwarfs}\label{sec:IntBrownie}
\subsection{Selection of a Brown Dwarf sample}\label{sec:BDsample}

The REOS.3 tables were built with the main purpose to calculate interior models and mass-radius relations (MRR) for BDs. 
Due to the lack of observational constraints on the internal structure other than mass and radius, we assume a BD to be a homogeneous sphere. Since there is an overlapping mass regime of BDs and GPs \citep{Hennebelle2008,Padoan2004}, we also assume that the GP KOI-889b consists of a homogeneous layer, in contrast to the three layer model applied for Jupiter, see the previous section.

The pressure-density profile of BDs and GPs is given by an adiabat that is fixed by an atmospheric boundary condition $P_{\mathrm{at}}$, see Sec.~\ref{subsec:Atmo}. The adiabat is derived from a linear mixture EOS with certain amounts of hydrogen (X), helium (Y) and heavier elements (Z). The interior profiles of an object with given mass $M$ and radius $R$ are derived by integrating the two following structure equations using a fourth-order Runge-Kutta scheme:
\begin{eqnarray}
 \frac{dr}{dm} & = & \frac{1}{4\pi r^2\rho} \;,\\
 \frac{dP}{dm} & = & -\frac{Gm}{4\pi r^4} \;.
\end{eqnarray}
$G$ is the gravitational constant and the Lagrangian coordinate $m$ is the mass within a shell of thickness $r$. The boundary conditions are given by $r(0)=0$, $r(M)=R$, and $P(M)=P_{at}$. 
The aim of these calculations is to study the influence of the EOS on the interior profiles by comparing results using the new REOS.3 tables and the SCvH EOS tables. 
Therefore, we do not consider the rather complex evolution of these objects with their characteristic early fusion processes, see \cite{Burrows2001}. 
For simplicity, our atmospheric profiles assume the same age of 5~Gyr for all objects studied here.

There are several BDs and a lot of massive GPs with measured mass and radius, see Fig.~\ref{fg:MRR}, from which we selected the GPs KOI-889b~\citep{Hebrard2013} and WASP-18b~\citep{Maxted2013} and the following BDs: Corot-3b~\citep{Deleuil2008}, Corot-15b~\citep{Bouchy2011a}, WASP-30b~\citep{Anderson2011}, KOI-205b~\citep{Diaz2013}, Kepler-39b~\citep{Bouchy2011b}, and LHS-6343c~\citep{Johnson2011}. We also plotted two predictions for the mass and radius of the first bona fide BD Gliese-229b~\citep{Nakajima1995} made by \cite{Marley1996} and \cite{Allard1996}.

In the following, we will study four of these objects with increasing mass in detail: KOI-889b, Corot-3b, Gliese-229b and Corot-15b. We use the input parameter mass $M$ in Jovian masses $\MJ$, radius $R$ in Jovian radii $\RJ$, and metallicity Fe/H of the host star from the literature as listed in Table~\ref{tab:objects}.

\begin{deluxetable}{cccc}[h!]
 \tablecolumns{2}
 \tablewidth{0pc}
 \tablecaption{Literature data of our studied objects.\\$^*$Note that mass and radius for Gliese-229b are derived from the fitting formulae given in \cite{Marley1996} while Fe/H = -0.2$\pm$0.4 is taken from \cite{Schiavon1997}. }
 \tablehead{ \colhead{Object} & \colhead{Mass ($\MJ$)}& \colhead{Radius ($\RJ$)} 
 & \colhead{Fe/H}   }
 \startdata
 KOI-889b & $9.98 \pm 0.5$ & $1.03 \pm 0.06$ & -0.07$\pm$0.15\\
 \citep{Hebrard2013}&&&\\
 Corot-3b & $21.66 \pm 1$ & $1.01 \pm 0.07$ & -0.02$\pm$0.06   \\
 \citep{Deleuil2008}&&&\\
  Corot-15b & $63.3 \pm 4.1$ & $1.12_{-0.15}^{+0.3}$ & 0.1$\pm$0.2   \\
 \citep{Bouchy2011a}&&&\\
  Gliese-229b$^*$ & $46.2_{-14.8}^{+11.8}$ & $0.87_{-0.07}^{+0.11}$ & -0.2$\pm$0.4   
 \enddata
 \label{tab:objects}
 \end{deluxetable}

\subsection{Atmospheric boundary conditions}\label{subsec:Atmo}

BDs and GPs host complex and actively circulating molecule-dominated atmospheres and, depending on the temperature, grains that lead to dusty atmospheres, 
see \cite{Burrows2011, Showman2013, Zhang2014}, and \cite{Basri2000,Baraffe2014} for reviews.

\begin{figure}
 \plotone{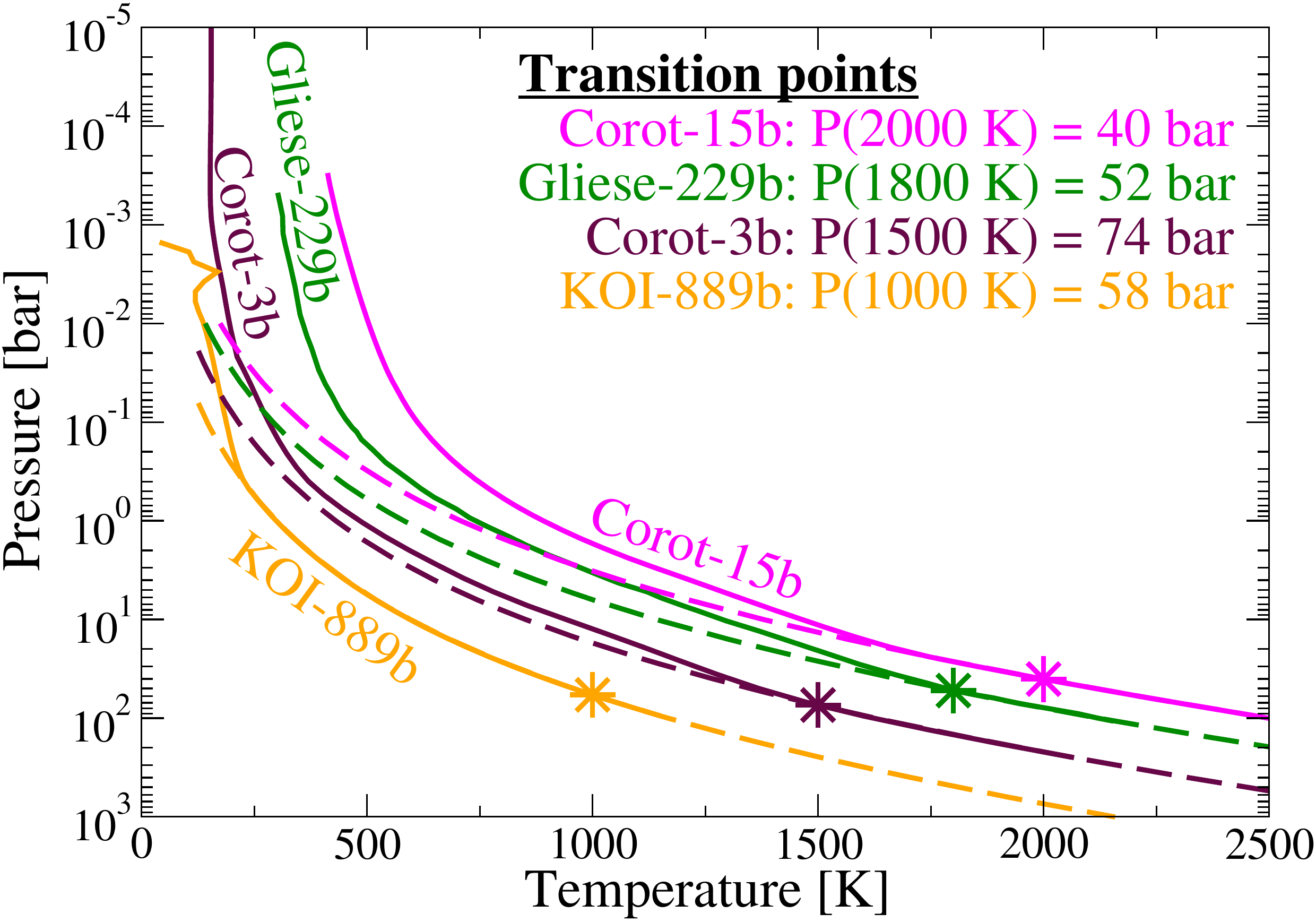}
 \caption{\label{fg:BDatm}  Solid lines: atmospheric profiles of KOI-889b (orange), Corot-3b (brown), Corot-15b (magenta) and Gliese-229b (green), for which data 
 was taken from \cite{Marley1996}. The dashed lines represent the respective adiabats. The stars indicate the transition points where the atmospheric profiles 
 becomes adiabatic ($P_{at}$) and, simultaneously, set the boundary condition for the interior profiles.}
 \end{figure}

Measured emitted spectra of the objects are generally necessary to obtain realistic atmospheric profiles, but evolutionary models can also be used to assess atmospheric 
structure.  For Gliese-229b, where accurate spectra have been obtained, but the mass and radius are poorly known, 
\citep{Allard1996, Marley1996, Oppenheimer1998, Saumon2000}, we use the atmospheric profile from \cite{Marley1996}, see Fig.~\ref{fg:BDatm}.
Since we focus on the effect of the EOS on the interior profile, we make some simplifying assumptions for the other objects KOI-889b, Corot-3b and Corot-15b. 
We neglect irradiation from the host star and set their age to 5~Gyr.  Given the measured masses we extracted the respective effective temperatures from the models of \cite{Baraffe2003}.  Based on the measured surface gravities and estimated effective temperatures for these objects, we used the radiative-convective atmosphere model of \cite{Marley1996} and \cite{Fortney2008b} to calculate the temperature structure, and the depth at which the radiative atmosphere becomes convective and stays convective.  This is the tie point for the deep atmospheric structure.

The resulting atmospheric profiles are shown as solid lines in Fig.~\ref{fg:BDatm}. We show the transition points, where the temperature profiles become adiabatic. These points define the atmospheric boundary condition $P_{\mathrm{at}}$ as well. We obtain for Corot-15b $P(2000~\textrm{K})=40$~bar, for Gliese-229b $P(1800~\textrm{K})=52$~bar, for Corot-3b $P(1500~\textrm{K})=74$~bar, and for KOI-889b $P(1000~\textrm{K})=58$~bar. The dashed lines in Fig.~\ref{fg:BDatm} represent the respective adiabats, which determine the interiors. While this is a simple prescription, since we are only interested in interior differences between EOS given reasonable upper boundary conditions, this treatment certainly suffices.

\subsection{Mass-radius relations}\label{sec:MRR}

\begin{figure}
 \plotone{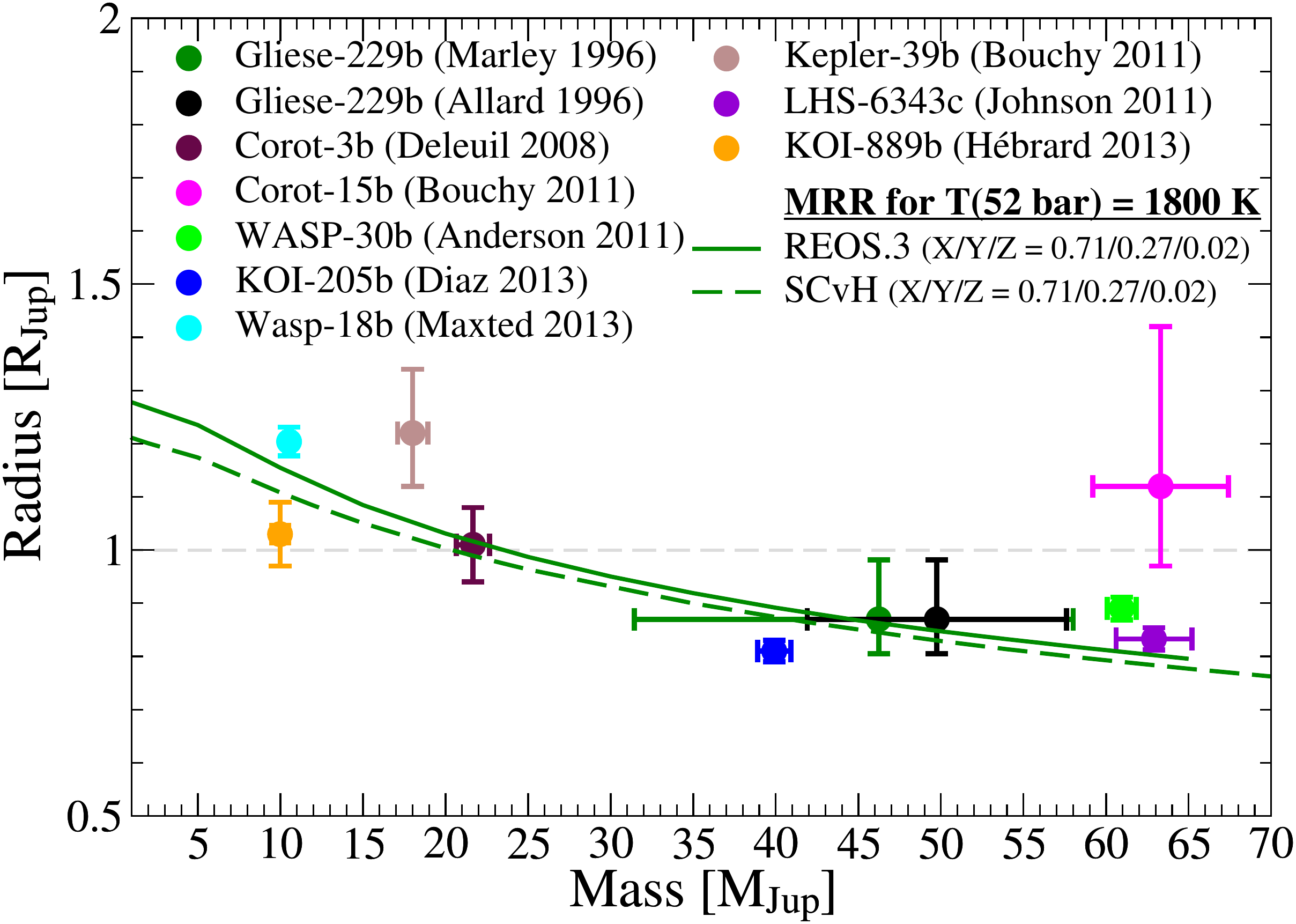}
 \caption{\label{fg:MRR} Map of selected BDs with known mass and radius and the Giant Planet KOI-889b. The solid green line shows a MRR for objects with solar-like composition $(X=0.71, Y=0.27, Z=0.02)$ and an atmospheric boundary condition of $T(52~\textrm{bar})=1800$~K using the REOS.3 . The respective SCvH-EOS result is represented by the dashed green line.}
\end{figure}

We have calculated MRR for homogeneous and adiabatic objects with an atmospheric boundary like Gliese-229b of $P(1800~\textrm{K})=52$~bar, a helium content of $Y=0.27$, and 
a metallicity of $Z=2\%$. The results for the different EOS are shown as green curves (solid: REOS.3, dashed: SCvH) in Fig.~\ref{fg:MRR}. We obtain systematically higher 
radii using the REOS.3 with a maximum deviation from the SCvH-EOS of $\sim6\%$ at $1~\MJ$. For masses above $20~\MJ$ the deviation remains nearly constant at $\sim2.5\%$. 
This is in agreement with \cite{Militzer2013b} who obtained slight radius enhancement using their \textit{ab initio} EOS data compared to the SCvH EOS for planets with 
masses between 0.5 and 2 Jovian masses. 
This finding can now be extended up to 70 Jovian masses and interpreted directly in terms of the EOS:

 The main deviations between REOS.3 and SCvH arise within the warm dense matter regime (WDM), where strong correlations and quantum effects are important and 
 dissociation and ionization processes occur. There, the DFT-MD EOS data are not as compressible as the SCvH ones, see the lower maximum compression at the principal 
 Hugoniot curve in Sec.~\ref{sec:intro}. Since the objects with lower masses around $1~\MJ$ are dominated by WDM, the more compressible SCvH data lead to significantly 
 smaller radii than the REOS.3 data. At higher densities and/or temperatures the system is fully ionized and becomes increasingly degenerate, which is accurately described by 
 both EOS leading to similar MRR.

The next aim was to investigate the variation of the radius with respect to the metallicity. For simplicity, the heavier elements ($Z$) are represented by a scaled helium 
EOS ($\rho_\mathrm{Z}(P,T)=4\times\rho_\mathrm{He}(P,T)$, $u_\mathrm{Z}(P,T)=u_\mathrm{He}(P,T)/4$). We increased the fraction of $Z$ within the mixture and 
calculated the respective radii for the mean value of the mass of our four selected objects and the maximum errors of these masses, see Tab.~\ref{tab:objects}. Each object has a different 
adiabat, fixed by the different atmospheric boundary, see Fig.~\ref{fg:BDatm}. 

\begin{figure}
 \includegraphics[scale = 0.35]{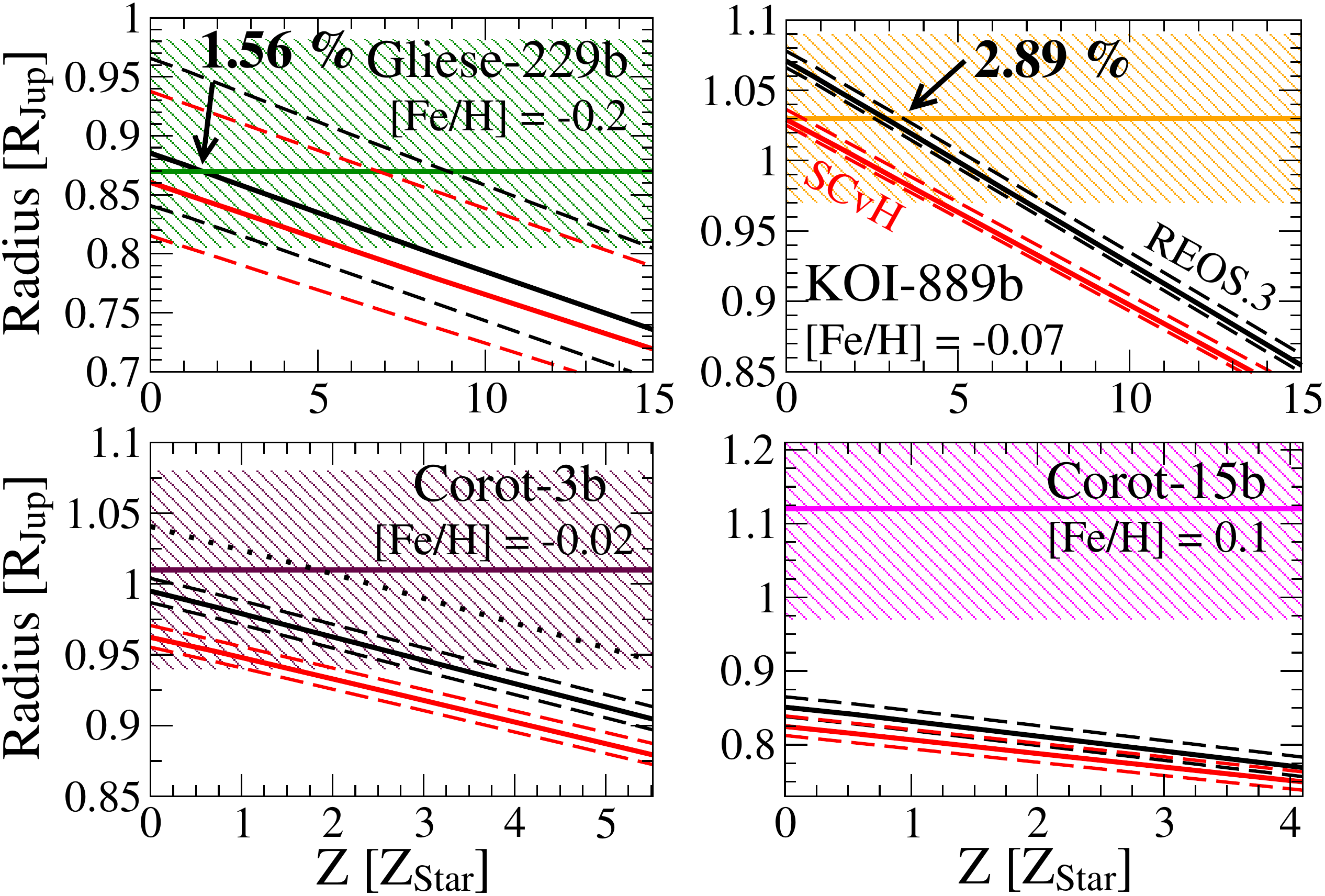}
  \caption{\label{fg:ZvsR} The radius of the BDs in dependence on the EOS and the metallicity with respect to the Fe/H ratio of the host star. The horizontal solid lines 
  indicate the mean values and the color-shaded areas the error bars of the respective observed radii, see Fig.~\ref{fg:MRR}. The solid black lines show the radii matching 
  the mean value and dashed black lines matching the error bars of the objects mass (see again Fig.~\ref{fg:MRR}) using the REOS.3 data. Red solid and dashed lines 
  represent the results using the SCvH EOS. The dotted line in the Corot-3b frame shows results assuming Gliese-229b-like adiabatic boundary condition 
  of $T(52~\textrm{bar})=1800$~K instead of our chosen $T(74~\textrm{bar})=1500$~K.}
\end{figure}

The results are shown in Fig.~\ref{fg:ZvsR}, where the metallicities are given in terms of the metallicity of the host star. The horizontal solid lines represent the 
mean values and the color-shaded areas indicate the error-bars of the objects observed radii.

The solid black lines represent our calculated radii with respect to the mean value of the mass using 
the REOS.3 data. Dashed black lines indicate the radii for the maximum observed uncertainties in total mass. The red solid and dashed lines illustrate the corresponding results 
using the SCvH-EOS.

Recall that for every object we try just one atmospheric boundary, meaning one tie point in $P-T$ for the interior adiabat.  This allows us to readily compare how 
different EOS affect the radius, while the remaining parameters are not changed.

In all cases we find a larger radius for the REOS.3, compared to SCvH.  For Gliese-229b the REOS.3 curve for the mean mass intersects the mean radius at a metallicity of $Z = 1.56~\%$. The SCvH radii have no value for the mean radius but are well located within its error bars. 
Due to the large error bars in mass, REOS.3 and SCvH results overlap and there are many possible metallicities up to $9~\%$ for the mean radius. 
For the other three objects, the accuracy of the metallicity derivation is of course tied to the simple choice of the atmospheric boundary condition, but the trends in radius with metallicity are clear.
In case of KOI-889b, we find disjoint solutions with respect to the underlying EOS. The REOS.3 result of $Z\sim3\%$ for the mean values of mass and radius is again slightly larger than the SCvH result.
The curves for Corot-3b are disjoint as well. Here the temperature dependence is nicely illustrated. For our chosen atmospheric conditions of $T(74~\mathrm{bar})=1500$~K we find no solution for the mean radius. But with a slightly higher initial temperature of $T(52~\mathrm{bar})=1800$~K, as for Gliese-229b, we obtain a metallicity of $\sim2\%$ with REOS.3 (dotted line in the Corot-3b frame).
In the case Corot-15b we fail to reach even the shaded error bar region of its radius with both EOS. This BD is very large for its mass compared to WASP-30b or 
LHS-6343c, see Fig.~\ref{fg:MRR}. Such a size can be explained by assuming a younger object age (since objects contract with time) or within advanced evolutionary calculations that assume a more opaque atmosphere slowing the cooling and contraction of the object, see \cite{Burrows2011}.

\subsection{Interior profiles}\label{sec:BDinterior}

\begin{figure}
\centering
  \plottwo{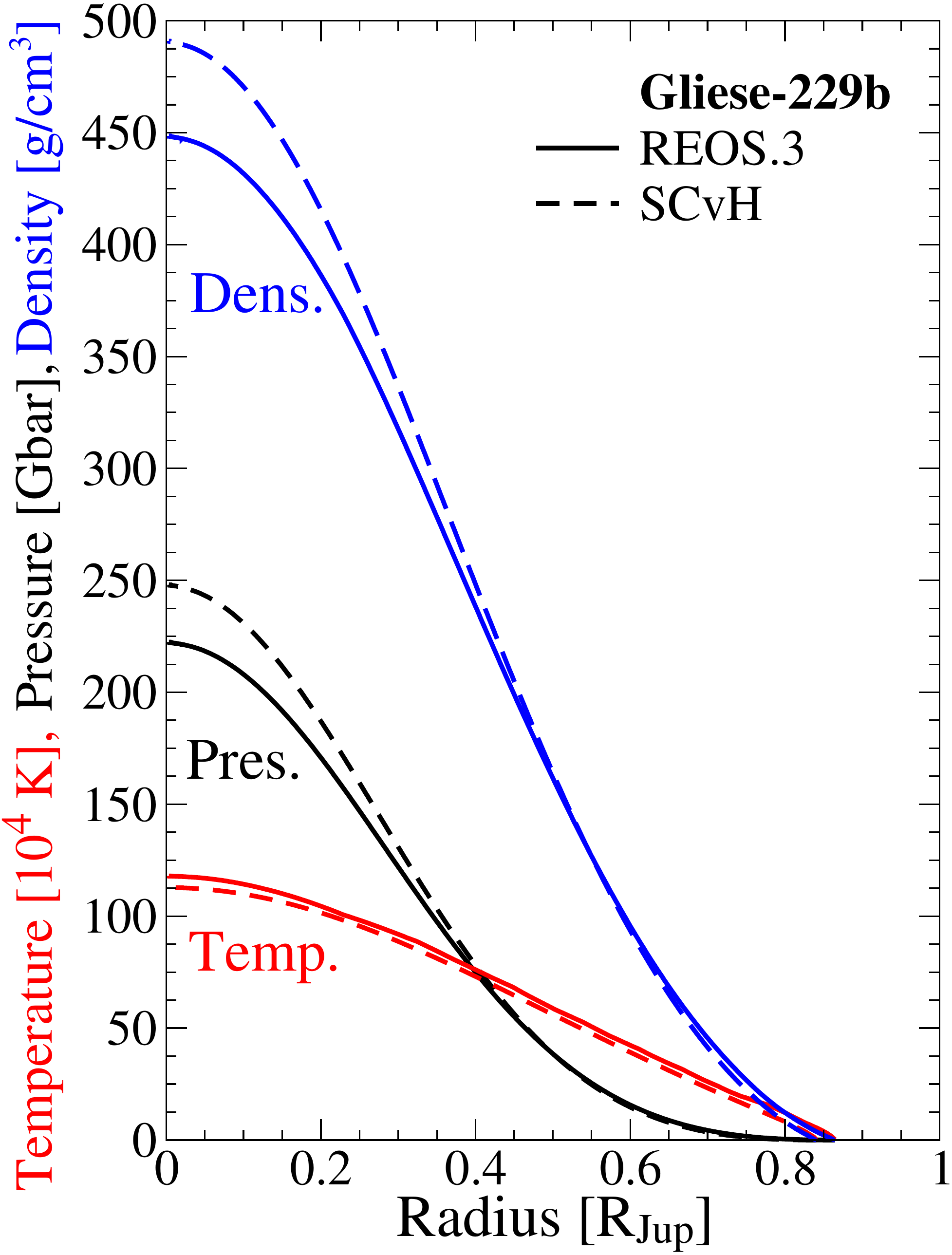}{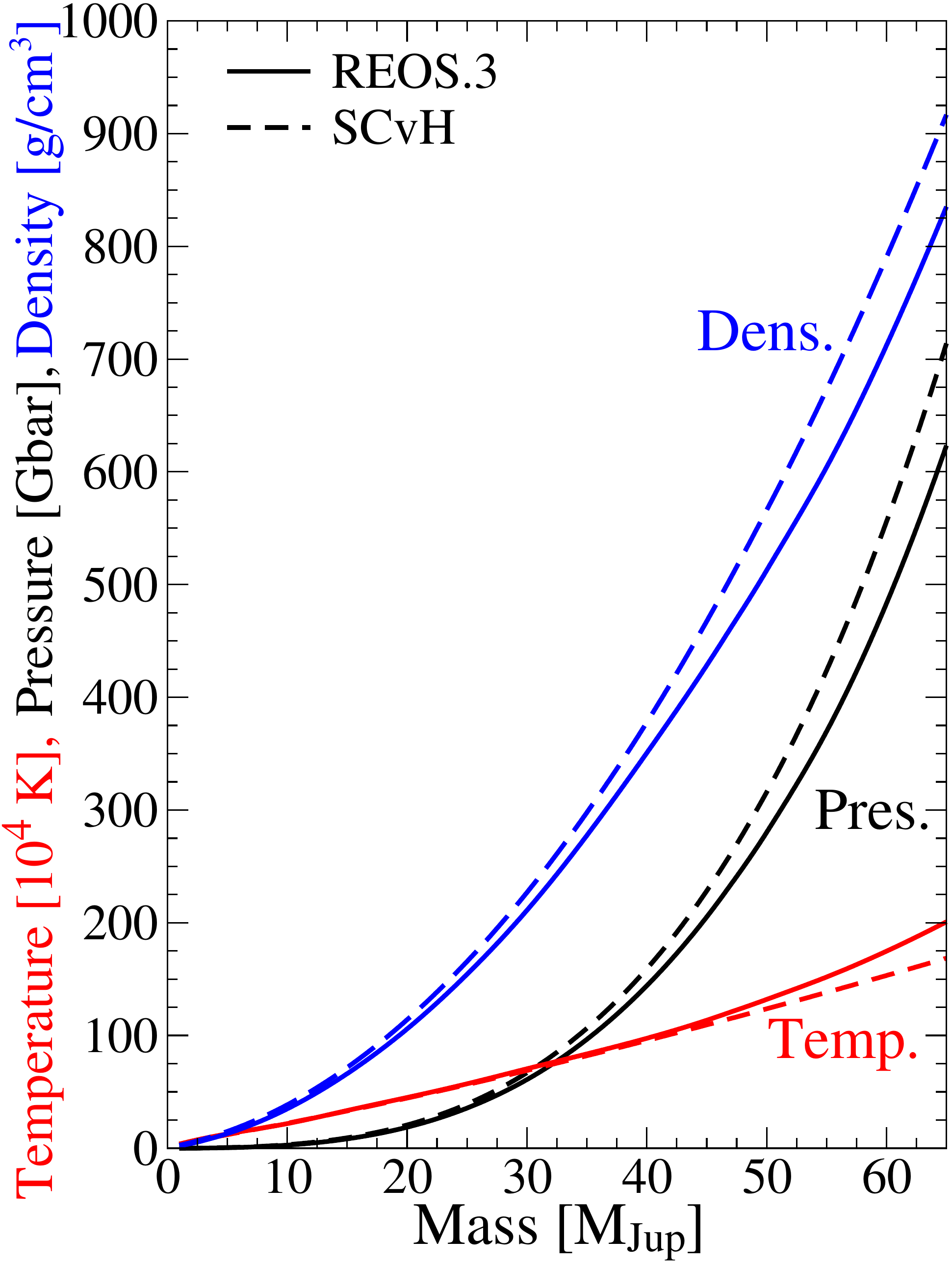}
 \caption{\label{fg:BDint1} Left panel: Interior profile of Gliese-229b using the SCvH-EOS (dashed) and the REOS.3 (solid). Right panel: Central conditions of objects with the same atmospheric conditions and the same composition like Gliese-229b depending on their mass. The appropriate mass radius relations are shown in Fig.~\ref{fg:MRR}.}
\end{figure}

\begin{figure}
 \plottwo{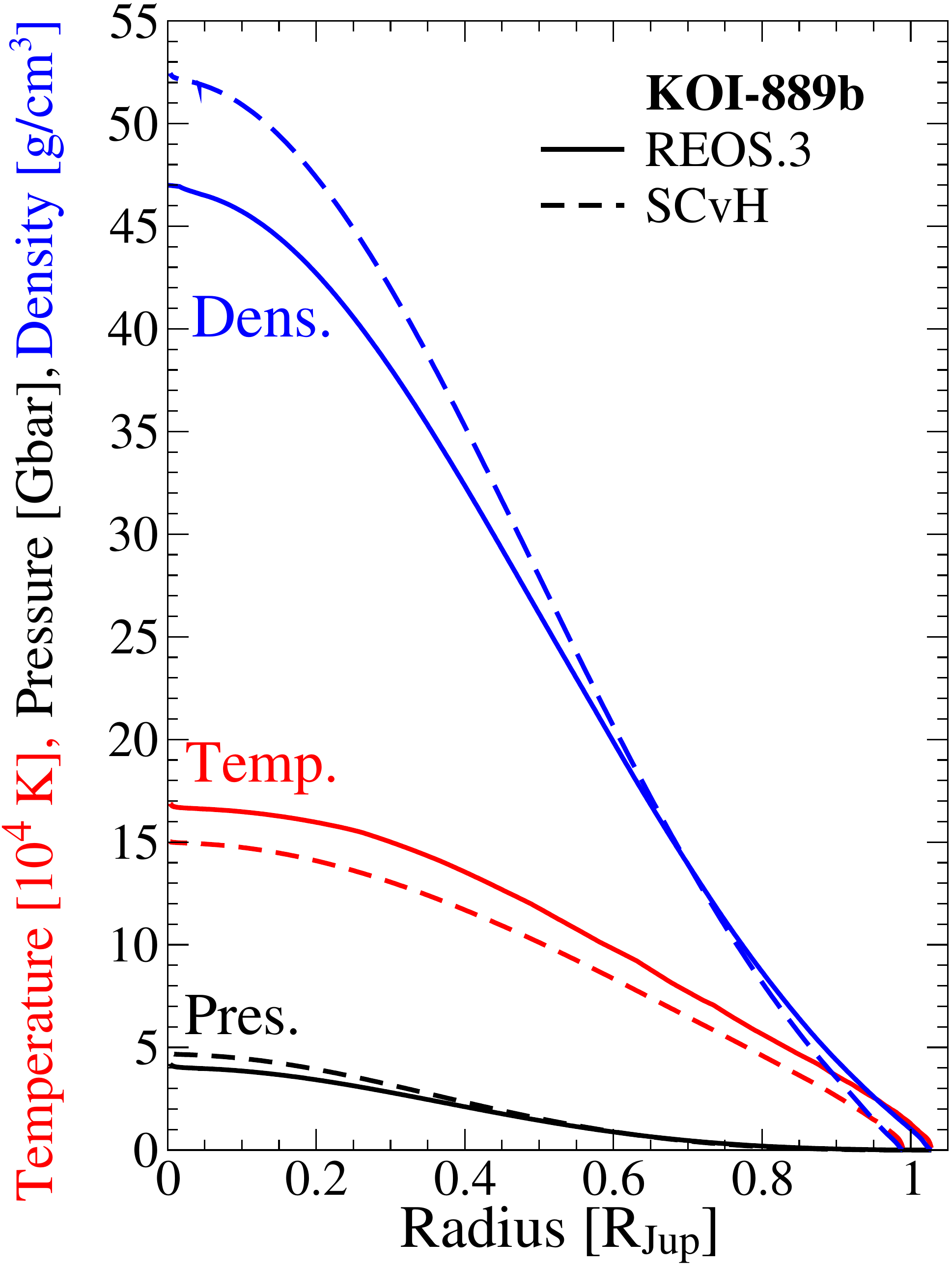}{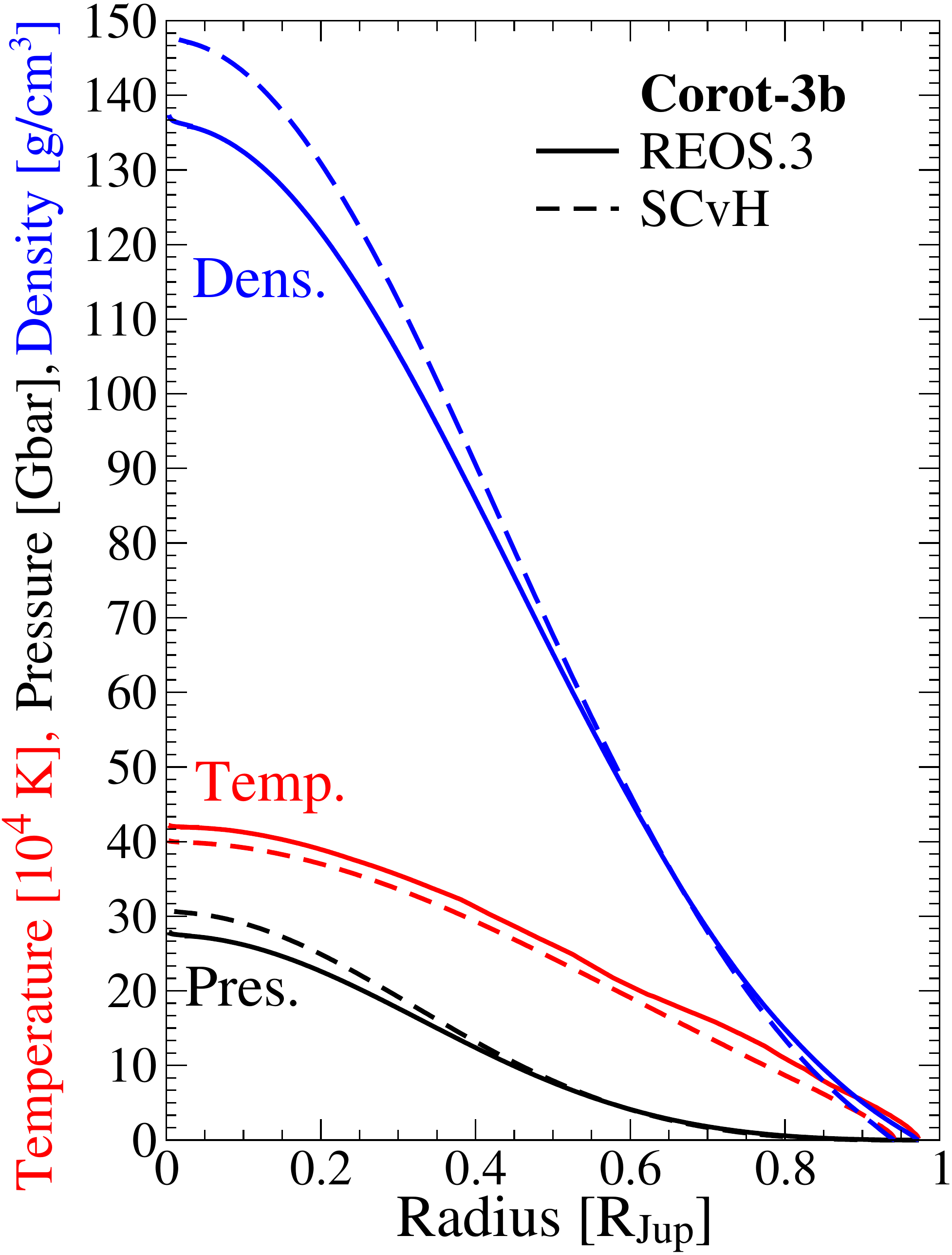}
 \caption{\label{fg:BDint} Interior profiles of KOI-889b \citep{Hebrard2013} (left panel) and Corot-3b \citep{Deleuil2008} (right panel) using the SCvH-EOS (dashed) and the REOS.3 (solid).}
 \end{figure}

Finally, we present interior profiles of Gliese-229b and Corot-3b calculated with $Z=2\%$ and of KOI-889b with a metallicity of $Z=4\%$ for the given masses and 
appropriate atmospheric conditions, see above. The results are shown in Figs.~\ref{fg:BDint1} and \ref{fg:BDint}. All these figures contain curves for the temperature 
in units of 10~kK (red), the pressure in Gbar (black) and the density using REOS.3 data (solid) and the SCvH EOS (dashed). The more massive an object, the higher are the central values, see right panel of Fig.~\ref{fg:BDint1} for objects with a composition and an atmospheric boundary like Gliese-229b. 
The respective MRR of these objects is shown in Fig.~\ref{fg:MRR}. 

While for the selected GP and BDs the central temperatures are similar but slightly higher using REOS.3, the central pressures of the SCvH results are higher by $\sim10\%$ due to the higher densities. Here one can see that the pressure is more influenced by the density than by the temperature as typical for degenerate matter. The higher pressures and densities of the SCvH results are a consequence of the predicted smaller radii, hence the objects are more compact.


\section{Conclusions}\label{sec:Conclusion}

We have constructed EOS tables for hydrogen and helium that include a large set of \textit{ab initio} data across the warm dense matter regime. 
Smooth connections to other EOS tables have been performed at those $T-\rho$ values where the DFT-MD technique as implemented in VASP either fails, due to the 
use of a plane wave basis set, or less expensive methods yield the same accurate results.

Compared to previous hydrogen EOS data from our group (H-REOS.1 and H-REOS.2), the new H-REOS.3 contains the same DFT-MD data set, with respect to the convergence criteria,
as is H-REOS.2 but is significantly extended to higher densities and temperatures. 

In contrast, the new He-REOS.3 is a substantial improvement for the entire $\rho-T$ plane.
It contains DFT-MD data derived from 108 particle (216 electrons) simulations for 21 isotherms between 60~K and 6~MK instead of 4 isotherms from 32-64 particle simulations 
within the former He-REOS.1. The ideal gas limit, the partially and fully ionized regime as well as the weakly coupled region below 10~kK are treated with 
sophisticated approaches. The final tables are thermodynamically consistent to a large extent.
Our tables are available within the online supplemental material of this paper.

We find a very good agreement of the theoretical principal Hugoniot curve calculated from our He-REOS.3 with the experiments of \cite{Nellis1984} that include EOS points 
in the vicinity of the low-pressure part of the Gliese-229b adiabat.

Furthermore, we find no significant deviations of the linear mixture of our EOS data from the real mixture results of \cite{Militzer2013b} in the regime relevant for BDs 
and, therefore, conclude that a linear mixing EOS suffices for interior models of BDs. Demixing effects do not occur in our investigated objects as well. For Jupiter, the new He-EOS (He-REOS.3) leads to slightly higher envelope metallicities with a maximum atmospheric enrichment of 3$\times$~solar while all previous results~\citep{Nettelmann2012} remain valid.

Mass-radius relations for the BD mass regime derived from our EOS data lead to higher radii (between $2.5-5\%$) and higher central 
pressures ($\sim10\%$) and densities ($\sim10\%$) compared to results using the SCvH EOS.

Further important issues in deriving highly-accurate EOS data for hydrogen and helium and modeling Giant Planets and Brown Dwarfs are, e.g., calculations for the 
free energy, or at least the entropy, within the H-REOS.3 and He-REOS.3, to construct a reasonable EOS covering a similar density-temperature range that represents heavier elements, e.g., by water, and to provide material properties along the adiabats of GPs and BDs as it was done by \cite{French2012} for Jupiter. This remains subject of forthcoming work.

\acknowledgments
We thank R.\ Hellmann, G.\ Chabrier, A.\ Potekhin, D.\ Saumon, T.\ D\"oppner, M.\ Bethkenhagen, M.\ French, and B.\ Holst for helpful discussions.
We kindly thank the anonymous referee for providing an insightful report.
This work was supported by the Deutsche Forschungsgemeinschaft within the SFB~652 and the SPP~1488. 
The DFT-MD simulations were performed at the North-German Supercomputing Alliance (HLRN) 
and at the IT- and Media Center of the University of Rostock.

\bibliographystyle{apj}
\bibliography{./WRBDarXiv}

\end{document}